\documentclass[lettersize,journal]{IEEEtran}
\usepackage{amsmath,amsfonts}
\usepackage{algorithmic}
\usepackage{algorithm}
\usepackage[table]{xcolor}
\usepackage{array}
\usepackage[caption=false,font=normalsize,labelfont=sf,textfont=sf]{subfig}
\usepackage{textcomp}
\usepackage{stfloats}
\usepackage{url}
\usepackage{makecell}
\usepackage[hidelinks]{hyperref} 
\usepackage{verbatim}
\usepackage{graphicx}
\usepackage{color}
\usepackage{flushend}
\usepackage{makecell}
\usepackage{cite}
\usepackage{enumitem}
\usepackage{multirow}
\usepackage{amssymb}
\usepackage{pifont}
\usepackage{bbding}
\begin{document}

\title{A Synergy of Computing Power Networks and Low-Altitude Economy Intelligent Communications: Challenges, Design Principles, and Research Directions}

\author{Yan Sun, Yinqiu Liu, Shaoyong Guo*, Ruichen Zhang, Jiacheng Wang, Xuesong Qiu,~\IEEEmembership{Senior Member,~IEEE,} Geng Sun, Weifeng Gong, Dusit Niyato,~\IEEEmembership{Fellow,~IEEE,} and Qihui Wu,~\IEEEmembership{Fellow,~IEEE}

\thanks{This work was supported by the National Natural Science Foundation of China (62322103). (*Corresponding author: Shaoyong Guo)}
\thanks{Yan Sun, Shaoyong Guo and Xuesong Qiu are with the State Key Laboratory of Networking and Switching Technology, Beijing University of Posts and Telecommunications, Beijing, 100876, China (E-mail: \{sunyan79, syguo, xsqiu, chenjiewei\}@bupt.edu.cn).}
\thanks{Yinqiu Liu, Ruichen Zhang, Jiacheng Wang and Dusit Niyato are with the College of Computing and Data Science, Nanyang Technological University, 639798, Singapore (E-mail: yinqiu001@e.ntu.edu.sg, ruichen.zhang@ntu.edu.sg, jcwang\_cq@foxmail.com, DNIYATO@ntu.edu.sg).}
\thanks{Geng Sun is with the College of Computer Science and Technology, Jilin University, Changchun 130012, China, and also with the College of Computing and Data Science, Nanyang Technological University, 639798, Singapore (email: sungeng@jlu.edu.cn).}
\thanks{Weifeng Gong is with the Inspur Computing Technology Pty Ltd, Beijing, 100095, China (E-mail: gongwf@inspur.com).}
\thanks{Qihui Wu is with the College of Electronic and Information Engineering, Nanjing University of Aeronautics and Astronautics, Nanjing 211106, China (e-mail: wuqihui@nuaa.edu.cn).}

}

\maketitle

\begin{abstract}
The rapid development of the Low-Altitude Economy (LAE) has created opportunities for emerging services such as autonomous aerial transportation, aerial sensing, and emergency response, all of which rely on efficient and intelligent communications. However, LAE intelligent communications face several challenges, including the limited computational capacity of aerial nodes, the lack of cross-scenario generalization, and the complexity of heterogeneous demands. Meanwhile, Computing Power Networks (CPNs) have emerged as a new paradigm for integrating distributed computing, networking, and storage resources, but they are also constrained by static deployment and limited adaptability. In this survey, we explore the synergy between LAE intelligent communications and CPNs. We first analyze how CPNs can support LAE intelligent communications in areas such as air–ground collaborative control, AI training, communication–computation co-ptimization, and ubiquitous low-altitude information processing. Conversely, we discuss how LAE intelligent communications can enhance CPNs through mobility-assisted control, distributed intelligent training, dynamic routing, and in-network aerial computing. Finally, based on these insights, we outline design principles and future research directions for integrated CPN–LAE systems. This work provides a comprehensive foundation for building flexible, adaptive, and resilient architectures that leverage the synergy between CPNs and LAE to deliver high-quality and sustainable low-altitude services.
\end{abstract}


\section{Introduction}
\subsection{Background}
With the rapid development of the Low-Altitude Economy (LAE), a wide range of services, such as Autonomous Aerial Vehicle (AAV) transportation, aerial sensing, logistics and parcel delivery, and emergency response, are being actively deployed to meet emerging societal and industrial demands \cite{ref5}. LAE is envisioned as a critical component of future intelligent transportation and service systems, enabling applications that integrate aerial mobility, real-time monitoring, and ubiquitous connectivity at low altitude airspace. This fast-growing ecosystem not only opens vast business opportunities but also raises urgent technical challenges related to scalability, safety, and efficiency\cite{ref111}. Within the construction of the LAE ecosystem, communication emerges as a primary issue\cite{ref20}. Unlike traditional terrestrial networks, LAE services involve highly dynamic topologies, frequent handovers, and stringent requirements for low-latency and high-reliability data exchange. For example, AAV-based transportation requires continuous high-throughput links to ensure navigation safety, while aerial sensing and emergency missions demand real-time information sharing across heterogeneous aerial and ground nodes. Without robust and intelligent communication support, the envisioned services of the LAE cannot achieve their intended Quality of Service (QoS) or guarantee user safety and operational stability\cite{ref6}. 

At present, AI-powered intelligent communications have become a key approach to enhancing the efficiency and stability of LAE networks, and their role is expanding rapidly with advances in artificial intelligence \cite{ref6}. LAE intelligent communications aim to solve two major challenges: the efficient utilization of limited spectrum and computational resources, and the dynamic establishment of communication links in highly mobile and heterogeneous environments. To this end, intelligent communication in LAE can be broadly categorized into four dimensions \cite{ref7}: classification intelligence, which identifies communication scenarios, user intents, and resource categories to enable targeted service provisioning; estimation and prediction intelligence, which leverages learning-based models to infer channel states, predict traffic demand, and anticipate mobility patterns for proactive optimization; generation intelligence, which synthesizes adaptive communication strategies, such as coding schemes or routing policies, in response to dynamic requirements; and decision-making intelligence, which performs task-driven resource allocation and network control under complex and uncertain environments. Together, these four types of intelligence form the foundation of adaptive and resilient communication in the LAE.
\begin{table*}[!t]
\centering %
    \centering
    \caption{Summary of Related Surveys}
    
    \renewcommand{\arrayrulewidth}{0.8pt} 
    \renewcommand{\tabcolsep}{10pt} 
    
    {\fontsize{8}{10}\selectfont 
     
    \begin{tabular}{m{0.8cm}|m{8cm}|m{1.2cm}|m{1.5cm}|m{1.2cm}|m{1.3cm}} 
        \hline
        \rowcolor{gray!20}
         \textbf{Ref.}  & \textbf{Overview} & \textbf{LAE} &\textbf{LAE intelligent communications} & \textbf{CPN} & \textbf{Synergy of LAE and CPN}\\ 
        \hline

         \cite{ref5}  &   A survey on relevant standards and core architectures for supporting the development of LAE networks. &\Checkmark&\XSolidBrush&\XSolidBrush&\XSolidBrush\\ 
        \cline{1-6}
        \cite{ref35}  &  A survey on framework that integrates technologies such as sensing, localization, and intelligent flight.  &\Checkmark&\XSolidBrush&\XSolidBrush&\XSolidBrush\\ 
        \cline{1-6}
         \cite{ref36}  &  A comprehensive overview of computation-intelligence-based networking and collaboration algorithms for LAE. &\Checkmark&\XSolidBrush&\XSolidBrush&\XSolidBrush\\ 
        \cline{1-6}
         \cite{ref37}  & A survey on satellite-enhanced architecture for the low-altitude economy and terrestrial networks, based on generative AI, LLM, and Agentic AI. &\Checkmark&\XSolidBrush&\XSolidBrush&\XSolidBrush\\ 
        \cline{1-6}
         \cite{ref38}  &  A comprehensive survey on various applications of RL in multi-AAV wireless communications. &\Checkmark&\Checkmark&\XSolidBrush&\XSolidBrush\\ 
        \cline{1-6}
          \cite{ref39}  &  A survey on LLM-based solution tailored for near-field communication scenarios in LAE networks. &\Checkmark&\Checkmark&\XSolidBrush&\XSolidBrush \\
          \cline{1-6}
          \cite{ref40}  &   A survey on RIS-assisted AAV networks and propose an energy-efficient AAV communication framework. &\Checkmark&\Checkmark&\XSolidBrush&\XSolidBrush\\ 
          \cline{1-6}
         \cite{ref4}  &   A survey on how CPNs interconnect ubiquitous heterogeneous computing resources through the network to enable unified orchestration and flexible resource allocation. &\XSolidBrush&\XSolidBrush&\Checkmark&\XSolidBrush\\ 
        \cline{1-6}
        \cite{ref12}   &   A survey on the definition, architecture, and advantages of CPN. &\XSolidBrush&\XSolidBrush&\Checkmark&\XSolidBrush\\ 
        \cline{1-6}
        \cite{ref13}  &   A survey on the research background, key technologies, and primary application scenarios of CPNs. &\XSolidBrush&\XSolidBrush&\Checkmark&\XSolidBrush\\ 
        \cline{1-6}
         Our survey &   A survey discussing the challenges of LAE intelligent communications and CPN and exploring how they collaborate to achieve better system performance.
 &\Checkmark&\Checkmark&\Checkmark&\Checkmark\\ 
          \cline{1-6}
    \end{tabular}}
\end{table*}

However, the development of LAE intelligent communications faces the following challenges:
\begin{itemize}
    \item \textbf{Limited computational capability of LAE nodes.} Due to energy consumption and flight efficiency constraints, LAE nodes cannot deploy high-performance computing hardware, resulting in limited capacity to handle intelligent tasks\cite{ref8}. How to meet the increasing demand for intelligence in large-scale communication environments is a critical challenge.
    \item \textbf{Lack of generalization in LAE communication intelligence.} Considering the mobility of LAE network nodes, they often need to handle communication tasks in various scenarios. However, due to their lightweight nature, they cannot deploy numerous intelligent models to cope with different conditions\cite{ref9}. How to efficiently utilize multi-scenario data and knowledge to train generalized intelligence for LAE communication remains an open problem.
    \item \textbf{Wide coverage and heterogeneous demands.} LAE networks extend communication coverage via mobile LAE nodes, but this also introduces issues such as massive data mixed transmission and diverse user communication requirements \cite{ref10,ref11}. How to efficiently transmit data and analyze heterogeneous user demands is another major challenge.
\end{itemize}

To address the above challenges, the Computing Power Network (CPN) provides an effective solution with its vision of ubiquitous computing and wide-area interconnection\cite{ref12}. Compared to the relatively isolated Mobile Edge Computing (MEC) and cloud computing architectures, CPN aims to integrate global computing resources and enable on-demand scheduling\cite{ref4}. Through information sharing and efficient collaboration among computing devices, CPN offers powerful global information perception and processing capabilities\cite{ref13}. For the limited computational capability of LAE nodes, CPN can leverage its ubiquitous computing resources to assist LAE nodes in processing intelligent communication tasks and dynamically allocate computing power across the network as needed\cite{ref12}. For the lack of generalization in LAE communication intelligence, CPN can utilize distributed computing to collect global data from LAE networks and transmit it to data centers. With their strong computing power, data centers can perform large-scale transfer learning and domain knowledge generalization based on this aggregated data\cite{ref14}. For the challenges introduced by wide coverage in LAE, CPN can rely on edge computing to accurately analyze data transmission requirements and user communication demands, and employ technologies such as Computing-aware Routing (CAR) and In-Network Computing (INC) to improve data transmission efficiency\cite{ref15,ref16}.

Conversely, LAE nodes, leveraging their mobility, can also assist CPN in achieving more efficient computing interconnection through intelligent communication functions, such as aerial computing resource discovery, computing-aware routing, and relay for computing traffic\cite{ref17,ref18}. Therefore, by integrating LAE intelligent communications with CPN and fostering their mutual collaboration, both LAE networks and CPN can be optimized, ultimately delivering higher-quality services for users.
\begin{table*}[!t]
\centering
\caption{Main Structure of Survey}
\renewcommand{\arrayrulewidth}{0.8pt}
\renewcommand{\tabcolsep}{10pt}
{\fontsize{8}{10}\selectfont
\begin{tabular}{m{2cm}|m{7cm}|m{7cm}}
    \hline
    \rowcolor{gray!20}
    \multicolumn{1}{c|}{} & 
    \multicolumn{1}{c|}{\textbf{Section III: CPN for LAE Intelligent Communications}} & 
    \multicolumn{1}{c}{\textbf{Section IV: LAE Intelligent Communications for CPN}} \\ 
    \hline
    Challenges & The challenges of LAE intelligent communications lie in limited computing power, training of general LAE intelligence, real-time processing of information, hybrid and complex routing and fixed network configuration. & CPN suffers from static and centralized control, limited adaptability to dynamic environments, inefficient communication-computing coordination and dynamic routing strategies.
     \\ 
    \hline
    \vspace{1.2cm}Solutions & \parbox[c]{7cm}{
    \begin{itemize}[leftmargin=*]
        \item Air-ground collaborative control [65], [67], [72], [73]
        \item CPN-assisted LAE intelligence training [44], [76], [79], [80], [81]
        \item Communication-computing collaborative optimization [82]-[84]
        \item Low-altitude routing optimization [88]-[91]
        \item Software-defined LAE intelligent communication [94]-[97]
        \item Ubiquitous processing of low-altitude information [100]-[105]
    \end{itemize}
    }
    & \parbox[c]{7cm}{
    \begin{itemize}[leftmargin=*]
        \item Air-ground collaborative control [115], [116], [119], [120]
        \item Air-ground collaborative intelligence training [123], [124], [126], [127]
        \item Communication-computing collaborative optimization [129], [130], [132], [133]
        \item Low-altitude routing optimization [135], [136], [138], [139]
        \item Software-defined LAE intelligent communication [141]-[144]
    \end{itemize}
    }
     \\ 
    \hline
\end{tabular}}
\end{table*}
\subsection{Related Surveys}
To clarify the differences between our work and related surveys, we analyze the unique contributions of our work in this section, as summarized in Table I.
\subsubsection{LAE Intelligent Communications}
Recently, LAE has attracted significant attention from researchers, and several related surveys have proposed various management and operational frameworks for LAE systems. For example, the authors in \cite{ref5} explored relevant standards and core architectures for supporting the development of LAE networks. They proposed design principles for integrating communication, sensing, computing, and airspace management technologies into the LAE network management framework, aiming to improve operational efficiency, optimize airspace utilization, and ensure safety. The authors in \cite{ref35} analyzed the challenges faced in dynamic airspace management, AAV operations, and security management in LAE, and proposed an LAE operational management framework that integrates technologies such as sensing, localization, and intelligent flight. This framework provides a fundamental reference for establishing a unified operational foundation and airspace management architecture for LAE. The authors in \cite{ref36} provided a comprehensive overview of computation-intelligence-based networking and collaboration algorithms for LAE, covering aspects such as network routing, cooperative task allocation, and collaborative path planning. The authors in \cite{ref37} analyzed and proposed a satellite-enhanced architecture for the low-altitude economy and terrestrial networks, based on generative AI, large language models (LLMs), and Agentic AI. 

As a fundamental component of LAE networks, intelligent communications have also attracted widespread attention. Researchers have conducted a series of surveys on how to enhance communication in LAE networks using AI. For example, the authors in \cite{ref38} summarized various applications of Reinforcement Learning (RL) in multi-AAV wireless communications, including data access, sensing and collection, resource allocation for wireless connectivity, and trajectory planning. The authors in \cite{ref39} analyzed the increased complexity of signal processing in LAE near-field communications and highlight the necessity of distinguishing between far-field and near-field users. They proposed an LLM-based solution tailored for near-field communication scenarios in LAE networks. The authors in \cite{ref40} provided a review of Reconfigurable Intelligent Surface (RIS)-assisted AAV networks and propose an energy-efficient AAV communication framework that integrates technologies such as trajectory optimization, power control, beamforming, and dynamic resource management.

\subsubsection{Computing Power Network}
CPN, as a novel network management paradigm, has attracted widespread attention for its potential to improve computing resource utilization and accelerate AI model training and inference. The authors in \cite{ref4} analyzed how CPNs interconnect ubiquitous heterogeneous computing resources through the network to enable unified orchestration and flexible resource allocation. They further elaborated on the fundamental architecture and key enabling technologies that support CPN functionality. The authors in \cite{ref12} provided a comprehensive overview of the definition, architecture, and advantages of CPN. They further elaborated on key issues such as computing power modeling, information perception, and network forwarding mechanisms. The authors in \cite{ref13} investigated the research background, key technologies, and primary application scenarios of CPNs. They analyzed how CPNs can meet the multi-level deployment and flexible orchestration requirements of computing, storage, and networking resources posed by future 6G services.

Considering the challenges in LAE intelligent communications we previously analyzed, CPN can effectively address these issues to enable more intelligent and efficient communications for LAE. On the other hand, LAE can also assist CPN in certain aspects. However, up to now, there is still a lack of a comprehensive survey that thoroughly discusses the synergy between them. In this paper, we fill this gap by providing an in-depth exploration of the design principles and research directions underlying this synergy.

\subsection{Our Contributions}
In this paper, we propose a synergy of CPN and LAE intelligent communications. To the best of our knowledge, this is the first work that systematically investigates the synergy between LAE intelligent communications and CPN, providing a unified perspective on their mutual enhancement. The main contributions are as follows:
\begin{itemize}
    \item We analyze the challenges of LAE intelligent communications and explore how CPN can address these issues. We categorize LAE communication intelligence into four types based on their application objectives. For each type, we identify its unique challenges and discuss the corresponding tasks in the physical layer, transmission layer, and application layer of LAE communication networks. Furthermore, we explore design principles for leveraging CPN to support these tasks and overcome the stated challenges.
    \item We explore design principles for leveraging LAE intelligent communications to assist CPN. To meet the demands of global computing resource awareness and interconnection in CPN, we propose design principles such as aerial computing-aware routing, aerial INC, providing scalable solutions to enhance CPN performance.
    \item We outline future research directions for integrating LAE intelligent communications with CPN, including digital twin, security and privacy protection, energy-saving, multi-agent coordination, space-air-ground integration, highlighting the significant potential of this synergy in improving computing resource utilization, communication efficiency, and service quality.
\end{itemize}

\subsection{Structure of This Paper}
The main structure of this paper is shown as Table II. Section II present an overview of LAE intelligent communications and CPN. Sections III to VI explore how CPN assist classification intelligence, estimation and prediction intelligence, generation intelligence and decision-making intelligence in LAE communications. Section VII explores design principles for leveraging LAE intelligent communications to assist CPN. Section VIII outlines the future directions. Finally, Section IX concludes this paper.

\section{Overview of LAE Intelligent Communications and Computing Power Network}
In this section, we first provide an overview of LAE intelligent communications and CPN, and then analyze the new capabilities that can emerge from their integration. Finally, we discuss the individual challenges faced by LAE as well as the new issues that arise from their combination. Fig. 1 illustrates the application scenarios of the integration between LAE and CPN.
\begin{figure*}[!t]
\centering
\includegraphics[width=\textwidth]{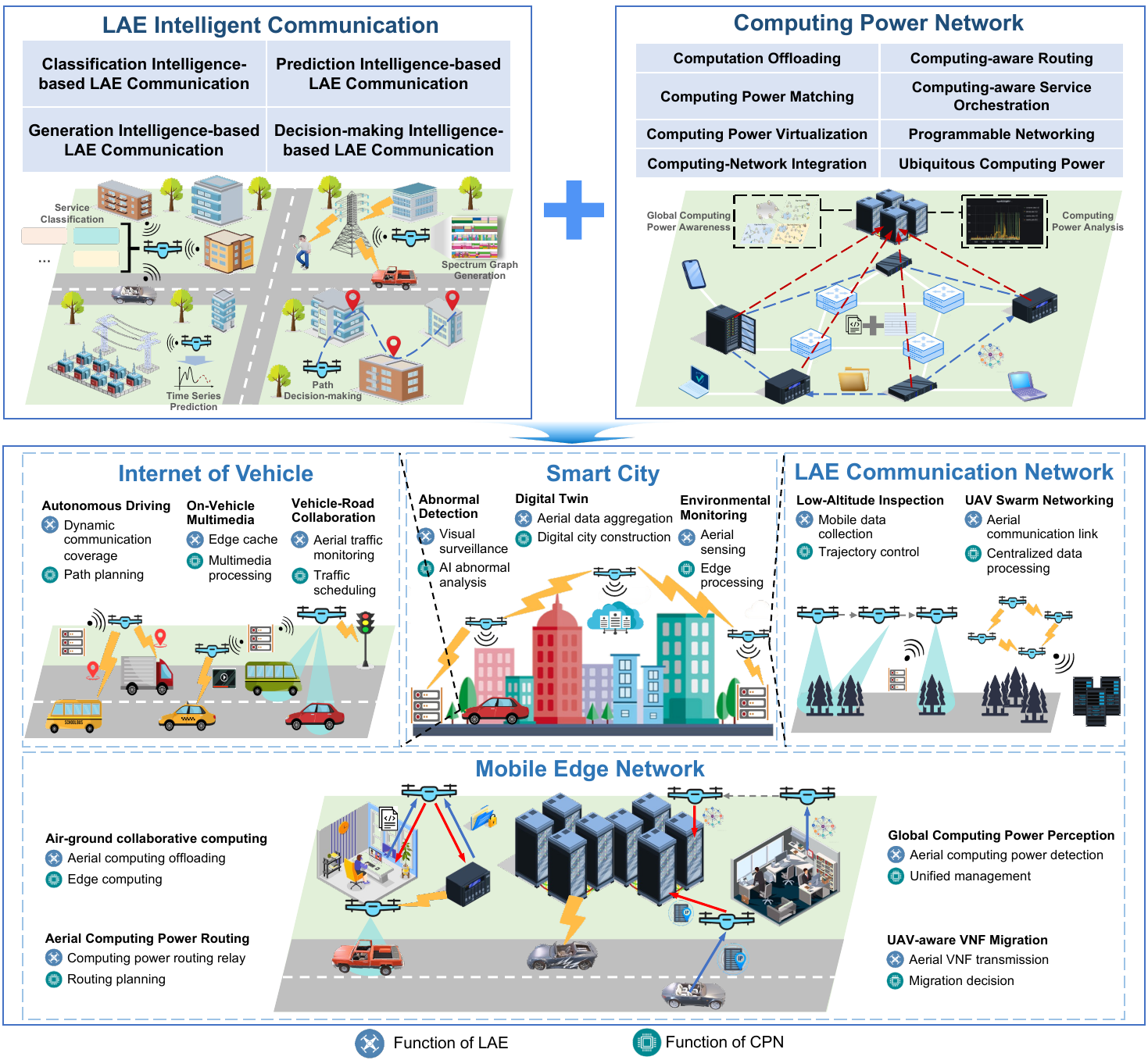}
\caption{Integrated framework of LAE intelligent communication and CPN. By integrating classification intelligence, predictive intelligence, generative intelligence, and decision intelligence in LAE communication with capabilities such as computation offloading and computing-aware routing in CPN, this framework can provide new capabilities across diverse scenarios. The figure shows the application scenarios of the integrated framework, including 4 scenarios and 12 use cases.}
\end{figure*}
\begin{table*}[!t]
\centering
\caption{Differences between LAE and AAV Technology}
\renewcommand{\arrayrulewidth}{0.8pt}
\renewcommand{\tabcolsep}{10pt}
{\fontsize{8}{10}\selectfont
\begin{tabular}{m{2cm}|m{7cm}|m{7cm}}
    \hline
    \rowcolor{gray!20}
    \multicolumn{1}{c|}{\textbf{Dimension}} & 
    \multicolumn{1}{c|}{\textbf{Autonomous Aerial Vehicle}} & 
    \multicolumn{1}{c}{\textbf{Low-Altitude Economy}} \\ 
    \hline
    Definition & A series of AAVs themselves and their control, perception, communication, and mission execution capabilities \cite{ref117}. & A series of emerging business models, aiming to fully exploit the airspace below 1,000 meters for a wide range of applications \cite{ref5}.
     \\ 
    \hline
    Coverage & The coverage of AAVs is generally limited by battery life, communication distance, and regulatory requirements, with a focus on specific operational scenarios. & LAE Faces the entire low-altitude airspace resources, it is necessary to coordinate airspace division, route planning, and air traffic control scheduling, covering from urban low altitude to cross regional low altitude transportation.
     \\ 
    \hline
    Supporting Technology & Flight control, power system, sensors, communication links, energy and endurance. & In addition to AAV related technologies, it also includes wireless communication (5G/6G), air traffic control system, edge computing, computing network, AI scheduling, low altitude surveillance radar. \\ 
    \hline
    Applications & Inspection, surveying, logistics, emergency rescue, communication relay. & Urban air traffic, low-altitude logistics, airspace management, communication network support, and service ecosystem construction.\\
    \hline
    Role in Industry & AAV technology provides hardware and control system products as execution terminals for LAE. & In addition to the tasks that AAVs are responsible for, LAE also includes service providers, regulatory platforms, communication and computing infrastructure, and so on. \\
    \hline
\end{tabular}}
\end{table*}
\subsection{Low-Altitude Economy}
With the rapid development of low-altitude flight technology, the AAV industry, and wireless communication technologies, low-altitude airspace is gradually becoming a new strategic resource following the ocean and outer space\cite{ref5}. Meanwhile, the growing societal demand for efficient logistics, smart city management, and personalized mobility is creating a vast market for LAE. Against this backdrop, LAE has emerged. LAE refers to a series of industrial activities and emerging business models centered around the utilization of low-altitude airspace, aiming to fully exploit the airspace below 1,000 meters for a wide range of applications such as logistics and transportation, emergency rescue, urban governance, and agricultural protection\cite{ref111}. This approach greatly reduces the operational burden on ground infrastructure while leveraging AAVs’ high mobility to provide ubiquitous, low-latency services. Consequently, LAE has become a research hotspot, giving rise to numerous studies addressing its existing challenges, such as low-altitude airspace resource planning and dynamic scheduling\cite{ref113}, privacy and information security protection\cite{ref112}, and low-altitude communication and navigation\cite{ref114}. These efforts are driving the sustainable and large-scale development of LAE.

As illustrated in Table III, compared to AAV networks that only focus on inter-AAV coordination, the LAE network is a comprehensive ecosystem-oriented framework designed for the entire low-altitude economy activities and transactions\cite{ref21}. By integrating technologies such as AI, communication, and Internet of Things (IoT), it delivers more comprehensive and intelligent services. For example, the authors in \cite{ref1} explored the management and resource allocation of low-altitude generative AI services. By leveraging AAVs' high mobility and low cost, their approach enables cooperation between multiple AAVs and ground edge devices to provide real-time generative AI computation services. Through a service management method aware of generative AI diversity and priority, this work achieves higher QoS. Similarly, the authors in \cite{ref111} proposed an air-ground integrated framework for low-altitude transportation using electric vertical take-off and landing (eVTOL) technology. This framework optimizes facility deployment and AAV fleet management, enhancing system performance, robustness, and autonomy in urban transportation. In summary, compared with UAV networks that are limited by factors such as battery life, communication distance, and regulatory constraints, the LAE network coordinates the entire low-altitude airspace, including airspace division, route planning, and air traffic control, thereby achieving broader and more systematic coverage. In terms of supporting technologies, while UAV networks mainly rely on flight control, sensors, and communication links, LAE networks additionally integrate 5G/6G wireless communication, edge computing, computing networks, and AI-based scheduling, enabling more intelligent and large-scale service orchestration.

\subsection{LAE Intelligent Communications}
As the foundation of LAE services, LAE communication underpins the operation of LAE networks. However, as the LAE network scales up and becomes increasingly dynamic, traditional rule-based or static communication schemes struggle to meet latency and reliability requirements. This is especially critical in airspace environments where resources are limited, or interference is significant\cite{ref20}. To address this, incorporating AI technologies to empower LAE communication with perception, reasoning, optimization, and learning capabilities has become a key research focus\cite{ref19}. Intelligent LAE communications enable real-time awareness of the communication environment in low-altitude airspace, accurate prediction of network states and service demands, and adaptive optimization of communication strategies. According to application objectives, intelligence for LAE communications can be broadly categorized into four types: classification intelligence\cite{ref19}, estimation and prediction intelligence\cite{ref22}, generation intelligence\cite{ref23}, and decision-making intelligence\cite{ref24}, as illustrated in Fig. 1 (a).

\subsubsection{Classification Intelligence} In LAE communications, classification intelligence is responsible for processing multi-dimensional data such as network state information, device operation status, and service requirements, and performing classification and discrimination. Specifically, it is widely applied in tasks like service type classification\cite{ref19}, demand classification\cite{ref25}, and anomaly detection\cite{ref26}. The classification results serve as the basis for further decision-making. Classification intelligence mainly relies on AI models such as Convolutional Neural Networks (CNNs), Graph Neural Networks (GNNs), and Support Vector Machines (SVMs)\cite{ref111}.

\subsubsection{Estimation and Prediction Intelligence} Considering the highly dynamic nature and uncertain links of LAE networks, achieving reliable, low-latency communication requires estimation and prediction intelligence to sense network conditions and forecast environmental changes\cite{ref5}. It is commonly applied in channel estimation, AAV trajectory prediction, and network traffic forecasting to support proactive network planning\cite{ref22,ref27}. Estimation and prediction intelligence is typically implemented using temporal AI models such as Long Short-Term Memory (LSTM), Gated Recurrent Unit (GRU), and Transformer.

\subsubsection{Generation Intelligence} Generation intelligence learns the feature distribution of original data to acquire the ability to generate new data. It can be applied to extension of LAE communication datasets or generate network configuration schemes for different scenarios\cite{ref23}. Furthermore, Transformer-based generation intelligence offers strong natural language understanding capabilities, enabling it to generate LAE network control instructions based on human language\cite{ref11}.

\subsubsection{Decision-Making Intelligence} To efficiently utilize communication resources and address challenges posed by high dynamism, LAE networks require decision-making intelligence that leverages the outputs of classification, prediction, and generation processes to make optimal decisions regarding resource allocation, link selection, and path planning\cite{ref24,ref28}. Decision-making intelligence is typically trained using Deep Reinforcement Learning (DRL) techniques.
\begin{table*}[!t]
\centering
\caption{Differences between CPN, edge computing network and cloud computing network}
\renewcommand{\arrayrulewidth}{0.8pt}
\renewcommand{\tabcolsep}{10pt}
{\fontsize{8}{10}\selectfont
\begin{tabular}{m{1.4cm}|m{4.6cm}|m{4.6cm}|m{4.6cm}}
    \hline
    \rowcolor{gray!20}
    \multicolumn{1}{c|}{\textbf{Dimension}} & 
    \multicolumn{1}{c|}{\textbf{Edge Computing Network}} & 
    \multicolumn{1}{c|}{\textbf{Cloud Computing Network}}&
    \multicolumn{1}{c}{\textbf{Computing Power Network}}\\ 
    \hline
    Definition & Edge computing uses network edge devices close to users to provide computing and storage capabilities, mainly addressing the needs of low latency, local processing, and real-time response \cite{ref118}. & Cloud computing relies on centralized data centers to provide large-scale, flexible computing power and storage, suitable for processing massive amounts of data and complex tasks\cite{ref119}. & CPN abstracts, schedules, and routes computing power resources from the cloud, edge, and end, emphasizing that computing power is a service, on-demand allocation, and global optimization \cite{ref12}.
     \\ 
    \hline
    Supporting Technology & Lightweight virtualization technology, 5G/6G access, multi access edge computing, privacy protection \cite{ref120}. & Backbone network, high-speed Data Center Interconnect (DCI), software-defined network, network function virtualization\cite{ref121}. & Programmable network, network slicing, in-network computing, computing-aware routing \cite{ref5,ref13}. \\ 
    \hline
    Applications & Intelligent manufacturing, autonomous driving XR, Real time control of AAVs\cite{ref122}. & AI model training, large-scale data analysis, video parsing, social platforms\cite{ref123}.& Cross domain AI services, cloud edge collaborative training, global resource optimization\cite{ref5}.\\
    \hline
    Characteristics & Low latency, local sensing, real-time control. & High computing power, centralized storage, large-scale. & Service perception, computing power scheduling, network programmability. \\
    \hline
    Deployment location & Base station, router, edge server. & Centralized large-scale data center. & CPN devices cover cloud, edge, and end, allowing services to be executed in the most suitable computing nodes.\\
    \hline
\end{tabular}}
\end{table*}
\subsection{Computing Power Networks}
CPN aims to interconnect computing, storage, and networking resources distributed across multiple layers, including cloud, edge and terminal, and construct a virtualized computing resource pool\cite{ref12}. It orchestrates and schedules computing resources based on service requirements to achieve global optimization of resource utilization. The differences between CPN, edge computing, and cloud computing are illustrated in Table IV. With the rapid development of AI, users' demand for computing infrastructure is increasing. Managing computing resources through CPN can significantly improve resource utilization, while efficient collaboration among computing facilities further enhances the execution efficiency of AI tasks\cite{ref13}. For example, the authors in \cite{ref3} proposed a generative AI service provisioning framework that integrates global computing resources, where large-scale models in cloud collaborate with lightweight edge models to provide generative AI services. They also introduced a globally coordinated AI training framework to deliver native generative AI services efficiently. Similarly, the authors in \cite{ref4} proposed a compute-first networking architecture that deeply integrates global computing resources using technologies such as Software-Defined Networking (SDN) and Network Function Virtualization (NFV). By combining centralized and distributed scheduling strategies, this architecture improves the utilization of computing resources. As shown in Fig. 1(b), the CPN possesses the following core capabilities \cite{ref4,ref12,ref13}:
\subsubsection{Computation Offloading} Computation offloading aims to intelligently migrate computing tasks originally executed locally to more suitable computing nodes, thereby enhancing the efficiency and performance of task execution. It involves comprehensive analysis of task characteristics, device capabilities, and network conditions to optimize the offloading strategy.
\subsubsection{Computing Power Matching} Computing power matching is designed to select the optimal computing resources from the global resource pool according to the computational requirements and QoS objectives of tasks. It takes task requirements and the collaboration potential into account to achieve optimal task-to-resource and resource-to-resource matching.
\subsubsection{Computing Power Virtualization} Computing power virtualization aims to uniformly encapsulate and manage heterogeneous computing resources such as CPU, GPU, and FPGA, forming a virtual computing resource pool. It abstracts the differences between diverse hardware infrastructures and provides standardized computing interfaces to users, while enabling on-demand resource allocation for different tasks.
\subsubsection{Computing-Network Integration} Based on technologies like In-Network Computing (INC), computing and networking capabilities can be deeply integrated, enabling the network not only to transmit data but also to perform certain data processing tasks, thereby reducing service response latency\cite{ref115}.
\subsubsection{Computing-aware Routing} CAR extends traditional routing decisions from being solely "optimization of transmission-optimal" to "co-optimization of computation and transmission." It considers not only conventional network metrics such as bandwidth and latency, but also the available computing power along the path and at the destination node, thereby dynamically routing tasks to the most suitable computing nodes.
\subsubsection{Computing-aware Service Orchestration} Computing-aware Service Orchestration enables the CPN to analyze users’ computing demands and QoS requirements, and, based on the real-time status of both network and computing resources, intelligently match tasks to computing nodes and dynamically schedule computing resources. This allows the CPN to deliver personalized and efficient computing services across diverse application scenarios.
\subsubsection{Programmable Networking} Programmable networking aims to enable dynamic customization and adjustment of task scheduling logic, transmission path selection, and resource allocation strategies within the CPN\cite{ref116}. Leveraging technologies such as SDN and Segment Routing over IPv6 (SRv6), CPN can flexibly adapt to diverse service requirements.
\subsubsection{Ubiquitous Computing Power} Ubiquitous computing power refers to the interconnection and integration of distributed edge and end computing resources and centralized cloud resources, allowing them to collaboratively execute tasks by leveraging their respective advantages.
\begin{figure*}[!t]
\centering
\includegraphics[width=\textwidth]{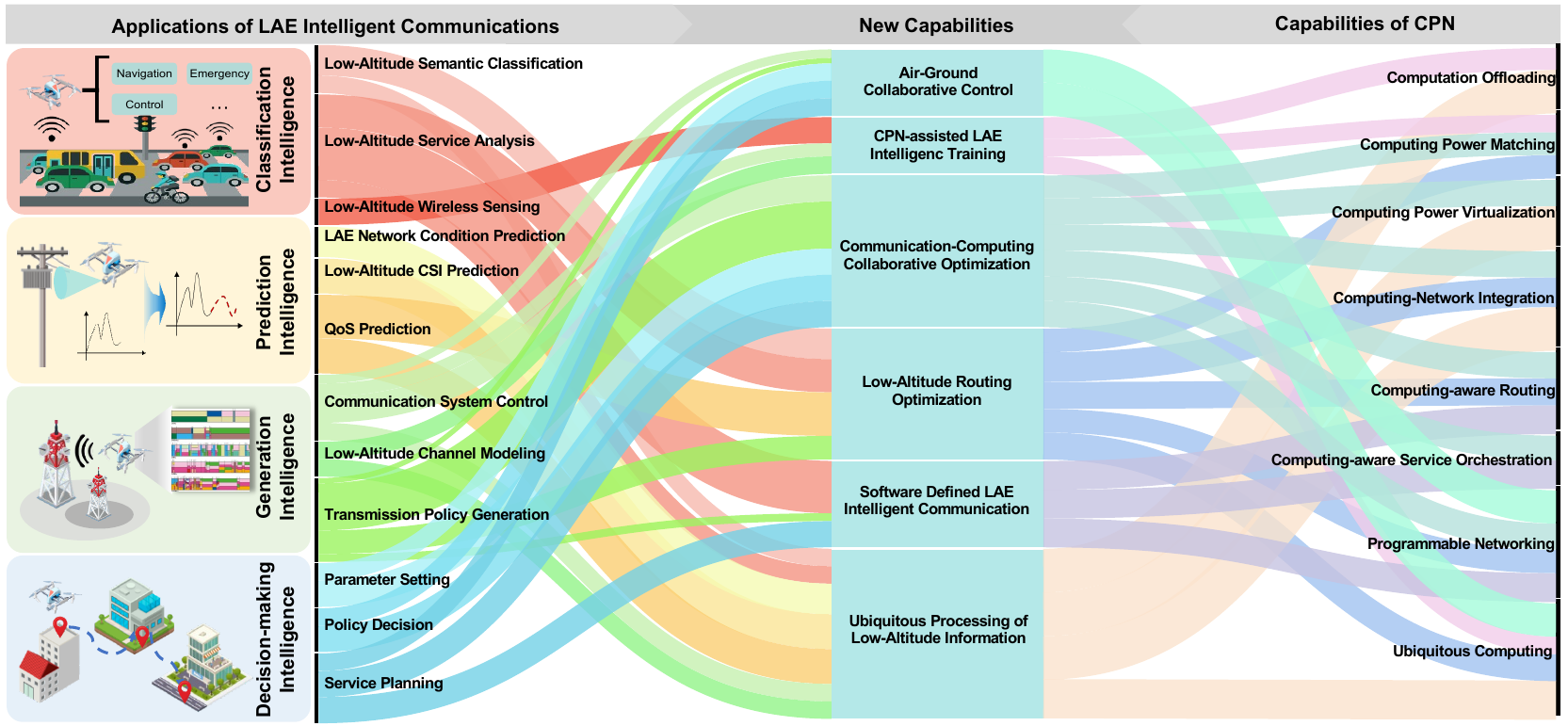}
\caption{The synergy of CPN and LAE intelligent communications. It brings new capabilities such as communication-computing collaborative optimization, ubiquitous processing of low-altitude information, low-altitude routing optimization, software defined LAE intelligent communication, air-ground collaborative control and CPN-assisted LAE intelligence training}
\end{figure*}
\subsection{Synergy of LAE Intelligent Communications and CPN: A Cloud-Edge-Aerial Node Integration Paradigm}
Based on the above review, we can observe that CPN and LAE intelligent communication can leverage their unique capabilities to assist each other. Specifically, for LAE intelligent communication, CPN can provide support in the following aspects:
\begin{itemize}
    \item \textbf{LAE service offloading}: Due to the limited computing power of LAE nodes, their service processing efficiency is relatively low. They can offload services to widely distributed computing power nodes to improve service execution efficiency \cite{ref68}.
    \item \textbf{Computing power matching of LAE service}: Considering that LAE nodes face diverse scenarios and that different services require different levels of computing power, CPN can quickly analyze service demands and match them with suitable computing resources \cite{ref69}.
    \item \textbf{Software-defined LAE communication network management}: With the support of CPN’s software-defined network management capabilities, LAE communication networks can offload management functions to CPN’s programmable control plane, enabling dynamic network management \cite{ref70}.
    \item \textbf{General LAE intelligent training}: The performance limitations of LAE nodes prevent them from deploying many AI models, which restricts their adaptability to diverse scenarios. CPN can use high-performance data centers to train general-purpose LAE intelligence to overcome this limitation \cite{ref50}.
\end{itemize}

On the other hand, for CPN, LAE can provide assistance in the following aspects:
\begin{itemize}
    \item \textbf{Reverse offloading of computing services}: During periods of service concurrency, computing nodes can offload part of the workload back to LAE nodes to reduce network congestion and latency. LAE nodes can also move toward areas with heavy computing demand to achieve better offloading effects \cite{ref71}.
    \item \textbf{Virtual Network Function (VNF) migration assistance}: With the high mobility and temporary computing assistance capabilities of LAE nodes, the VNF in CPN can be migrated via LAE nodes when network congestion, node overload, or task hotspot migration occurs \cite{ref67}. On the one hand, VNF can first be offloaded to nearby LAE nodes to maintain service continuity. On the other hand, LAE nodes can also serve as computing relays to smoothly migrate VNF to more suitable computing power nodes.
    \item \textbf{Expansion of CPN coverage}: By leveraging their mobility, LAE nodes can help remote users or computing devices access CPN, thereby extending its coverage \cite{ref72}. In addition, LAE nodes can detect computing devices which are not yet connected to CPN during inspection, assisting CPN in achieving global computing power awareness.
    \item \textbf{Dynamic computing power routing}: LAE nodes can act as relay in computing-aware routing, transmitting data to the appropriate computing devices. They can dynamically converge in service hotspot areas to reduce transmission delay \cite{ref73}.
    \item \textbf{Aerial in-network computing}: LAE nodes can be equipped with in-network computing hardware such as Data Processing Units (DPUs), enabling them to preprocess data while transmitting computing services, thus reducing service execution latency \cite{ref74}.
\end{itemize}

In summary, the synergy between LAE intelligent communication and CPN generates a new cloud-edge-aerial node integration paradigm, where the cloud and edge refer to CPN nodes and aerial nodes refer to AAVs. The synergy can generate new capabilities, as illustrated in Fig. 2.
\subsubsection{Air-Ground Collaborative Control} The control aims to achieve globally optimized control of LAE intelligent communication tasks through collaborative sensing and decision-making between LAE nodes and the ground computing nodes \cite{ref75}. The widely distributed ground computing nodes in the CPN can perceive the status of the entire LAE network in real time and, in combination with the sensing information from LAE nodes, undertake part of the intelligent communication task orchestration.
\subsubsection{CPN-assisted LAE Intelligence Training} The training aims to leverage the data from LAE communication networks within the respective regions of distributed computing nodes in the CPN to train localized LAE communication intelligence\cite{ref45}. At the same time, it reuses heterogeneous data and models, utilizing high-performance data centers for model generalization and transfer learning, thereby enabling the development of generalized LAE communication intelligence adaptable to diverse scenarios.
\subsubsection{Communication-Computing Collaborative Optimization} The optimization aims to jointly consider both the communication and computing requirements of LAE intelligent communication tasks, and dynamically adjust communication policies based on the low-altitude communication network and CPN environment\cite{ref44}. Some computation-intensive tasks such as path planning and channel prediction are offloaded to computing nodes located near the communication path. By collaboratively optimizing communication link selection and computing resource allocation, LAE intelligent communication can minimize task latency and ensure quality of service.
\subsubsection{Low-Aititude Routing Optimization} The routing optimization aims to leverage the computing-aware routing mechanism in the CPN to select the most suitable computing nodes and transmission paths for LAE intelligent communication tasks with diverse computational requirements. By utilizing the global computing awareness and programmable transmission capabilities of computing-aware routing, the system can dynamically generate transmission paths for LAE intelligent communication tasks based on task type, link state, and computing demands\cite{ref15}.
\subsubsection{Software Defined LAE Intelligent Communication} The software defined LAE intelligent communication aims to leverage the programmable control and centralized orchestration capabilities of the CPN to achieve dynamic management and on-demand optimization of LAE intelligent communication systems. Through the CPN control plane, the system can dynamically configure communication protocols, spectrum allocation, and computation offloading paths. Meanwhile, the service-aware and globally aware features provided by CPN enable flexible orchestration and migration of LAE intelligent communication tasks\cite{ref47}.

\subsubsection{Ubiquitous Processing of Low-Altitude Information} This function aims to process the massive, multi-source heterogeneous data generated by large-scale LAE nodes through ubiquitous computing resources, enabling efficient information perception and collaborative processing. Multimodal data such as visual, radar, and remote sensing signals—from LAE nodes are preprocessed in real time at the edge or terminal-side computing resources\cite{ref46}. By leveraging the computing interconnectivity of the CPN, it integrates information from multiple low-altitude and ground nodes to achieve situational awareness and collective intelligent decision-making.
\begin{figure*}[!t]
\centering
\includegraphics[width=\textwidth]{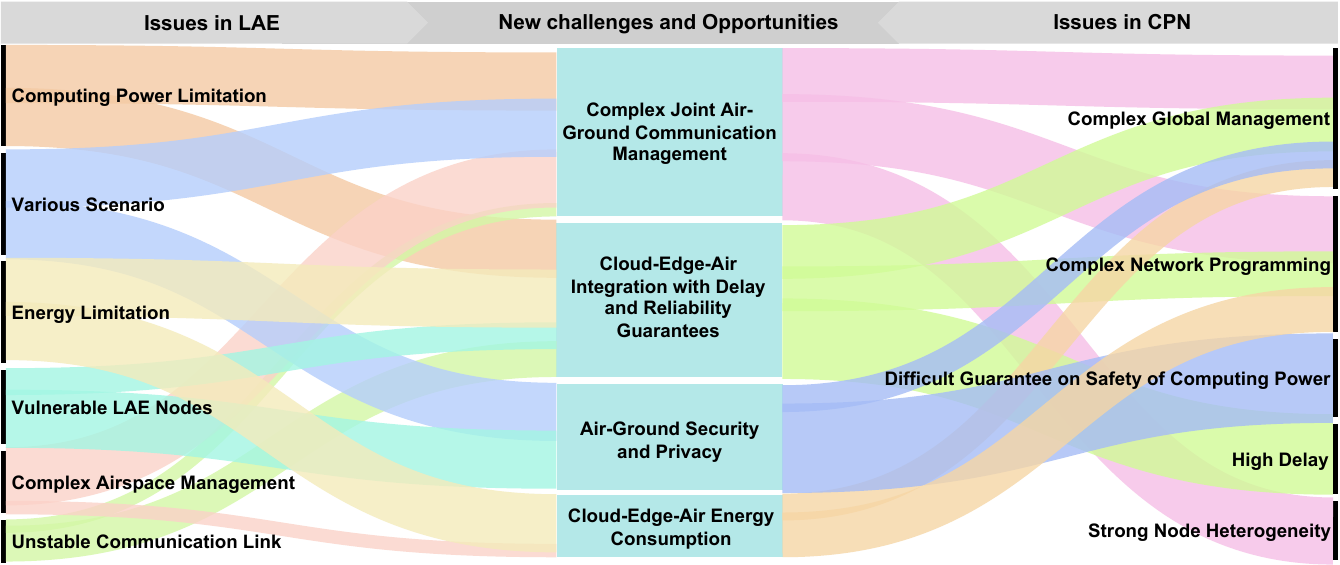}
\caption{The issues of CPN and LAE intelligent communications. The synergy brings joint issues such as complex air-ground joint management, difficult guarantee on communication delay and reliability, significant energy consumption and difficult guarantee on safety and privacy}
\end{figure*}
\subsection{Challenges in Integrated CPN and LAE Intelligent Communications}
Although the synergy between CPN and LAE intelligent communication can generate many new capabilities, considering that both of them inherently have issues to be addressed, their integration will also give rise to new joint issues as illustrated in Fig. 3.
\subsubsection{Complex Joint Air-Ground Communication Management} CPN itself faces the challenges of highly heterogeneous nodes and difficulties in global computing power management\cite{ref64}. After integration with LAE, the problem of insufficient computing power at LAE nodes needs to be compensated by CPN. Meanwhile, since LAE nodes operate in diverse scenarios, how to match them with suitable computing nodes becomes a problem to be solved. These make global computing power management more complex. In addition, the LAE communication network has a wide coverage and complex operation, making air-ground joint management for maximizing network performance another critical challenge. Facing this challenge, some AAV-ground joint optimization methods were proposed. For example, the authors in \cite{ref76} proposed a multi-objective optimization framework for joint AAV-AGV collaborative beamforming, leveraging extended multiobjective ant-lion optimization method to jointly optimize the location of AAVs and data transmission of automated guided vehicle.
\subsubsection{Cloud-Edge-Air Integration with Delay and Reliability Guarantees} Although LAE can assist CPN in data transmission and coverage expansion, low-altitude communication links are unstable and channels are easily affected by interference\cite{ref65}. At the same time, computing services in CPN, such as AI training, involve large data volumes and high computing delays. These issues together make it difficult to guarantee communication delay and reliability when LAE and CPN are combined. Facing this challenge, some AAV-ground communication delay and reliability guarantee methods were proposed. For example, the authors in \cite{ref77} proposed a joint AAV deployment and computation offloading method to minimize the response delay in AAV-assisted MEC system and improve the reliability of communications between AAV and edge devices.
\subsubsection{Air-Ground Security and Privacy} CPN aims to improve service execution through multi-node collaboration. However, such collaboration increases the risks of network attacks and privacy leakage\cite{ref66}. Considering the weak computing power and vulnerability of LAE nodes, the integration with LAE will further exacerbate the security risks of CPN. Facing this challenge, some joint AAV-ground security and privacy protection methods were proposed. For example, the authors in \cite{ref78} proposed a secure AAV delivery methods based on differential privacy and diffusion models to avoid privacy leak in the interaction between AAV and edge devices.
\subsubsection{Cloud-Edge-Air Energy Consumption} After integration with CPN, LAE nodes not only need to maintain their original flight and communication tasks but also take on part of the computing tasks. Moreover, air-ground joint management will induce more frequent movement of LAE nodes to meet CPN requirements, which leads to increased energy consumption. Facing this challenge, some low-consumption AAV-ground cooperation mechanisms were proposed. For example, the authors in \cite{ref79} proposed a successive convex approximation algorithm to jointly optimize the edge computing offloading and AAV trajectory, minimizing the energy consumption of AAVs and edge devices.

In summary, given the new capabilities and challenges arising from the synergy between LAE and CPN, how to design a efficient and effective interaction mechanism to maximize the benefits of their synergy while mitigating the impacts of joint issues is a key problem to be solved.
\begin{table*}[!t]
\centering
\caption{Summary of Related Works in CPN for LAE Intelligent Communications}
\renewcommand{\arrayrulewidth}{0.8pt}
\renewcommand{\tabcolsep}{10pt}
{\fontsize{8}{10}\selectfont
\begin{tabular}{m{2cm}|m{2cm}|m{12cm}}
    \hline
    \rowcolor{gray!20}
    \multicolumn{1}{c|}{\textbf{Topic}} & 
    \multicolumn{1}{c|}{\textbf{References}} & 
    \multicolumn{1}{c}{\parbox{12cm}{\textbf{Challenges of the LAE intelligent communications, techniques and advatages of proposed methods}}} \\ 
    \hline
    Air-ground collaborative control & [65], [67], [72], [73] & \makecell[l]{$\times$ \textbf{Challenge}: The limited computing power of LAE nodes results in suboptimal  performance of LAE\\ network control. \\ \checkmark \, \textbf{Techniques}: ubiquitous computing power, programmable networking, service migration. \\ \checkmark \, \textbf{Advantage}: \, CPN provides global computing power and programmable network capabilities, enabling\\ LAE nodes to generate more accurate communication strategies and reduce decision latency.}  
     \\ 
    \hline
    CPN-assisted LAE intelligence training & [44], [76], [79], [80], [81] & \makecell[l]{$\times$ \textbf{Challenge}: LAE nodes are lightweight and cannot deploy multiple models, making it difficult to adapt\\ to multiple scenarios. \\ \checkmark \, \textbf{Techniques}: \, Distributed Training, Transfer Learning, Digital Twin. \\ \checkmark \, \textbf{Advantage}: CPN utilizes distributed and high-performance computing centers to achieve model\\ migration, generalization, and digital twin verification, improving the efficiency and adaptability of LAE\\ intelligence.}
     \\ 
    \hline
    Communication-computing collaborative optimization & [82], [83], [84] & \makecell[l]{$\times$ \textbf{Challenge}: LAE nodes are difficult to meet both communication and computing needs, resulting in \\low efficiency in task offloading and resource scheduling. \\ \checkmark \, \textbf{Techniques}: \, Dynamic resource scheduling, task offloading, integrated sensing, communication and\\ computing \\ \checkmark \, \textbf{Advantage}: CPN supports communication perception, task offloading, and multi AAV collaboration to\\ achieve joint optimization of communication and computing, ensuring QoS for low altitude services.} \\
    \hline
    Low-altitude routing optimization & [88], [89], [90], [91] & \makecell[l]{$\times$ \textbf{Challenge}: Multiple types of data are mixed in large-scale low altitude transmission, making it\\ difficult for LAE nodes to accurately match computing power nodes. \\ \checkmark \, \textbf{Techniques}: \, Computing-aware routing, SDN, programmable routing. \\ \checkmark \, \textbf{Advantage}: CPN combines AAV maneuverability and computation aware routing to dynamically\\ construct integrated air ground paths, enhancing routing flexibility and reliability.}\\
    \hline
    Software defined LAE intelligent communication & [94], [95], [96], [97] & \makecell[l]{$\times$ \textbf{Challenge}: LAE nodes need to handle multiple businesses simultaneously and lack cross layer\\ management capabilities, making it difficult to design differentiated strategies for different businesses. \\ \checkmark \, \textbf{Techniques}: \, Computing-aware service orchestration, network function virtualization. \\ \checkmark \, \textbf{Advantage}: Relying on the programmable and centralized control capabilities of CPN, LAE\\ communication networks can achieve flexible strategy orchestration and adaptive management.}\\
    \hline
    Ubiquitous processing of low-altitude information & [100], [101], [102], [103], [104], [105] & \makecell[l]{$\times$ \textbf{Challenge}: Massive multimodal data results in high processing latency for LAE nodes and makes it\\ difficult to achieve global collaborative perception. \\ \checkmark \, \textbf{Techniques}: \, Compute offloading, edge/cloud collaborative processing, in network computing \\ \checkmark \, \textbf{Advantage}: CPN integrates multi-source heterogeneous computing power to perform real-time \\preprocessing and collaborative analysis on massive low altitude perception data, achieving global \\situational awareness and intelligent decision-making.}\\
    \hline
\end{tabular}}
\end{table*}
\section{CPN for LAE Intelligent Communications}
With the synergy between LAE and CPN, the novel capabilities enabled by CPN’s core functions can effectively tackle many challenges in LAE intelligent communications. In this section, we present a detailed analysis of these capabilities, as illustrated in Fig. 2, highlighting how CPN supports their implementation. We summarize the related works in Table V.
\subsection{Air-Ground Collaborative Control}
As shown in Fig. 2, air-ground collaborative control mainly involves generation and decision-making intelligence tasks in LAE communications. In this regard, generation intelligence is primarily responsible for generating LAE communication policies and controlling communication systems based on natural language. Generative models, represented by diffusion models and Generative Adversarial Networks (GANs), demonstrate strong robustness and exploration capabilities, achieving effective control strategy generation in complex and diverse LAE scenarios\cite{ref82,ref83}. Meanwhile, transformer-based generative models possess powerful natural language understanding and content generation capabilities, enabling user-centric control of LAE communication networks\cite{ref84}. However, the limited computing power of LAE nodes results in suboptimal performance of generation intelligence in producing communication policies\cite{ref80}, while the diversity of scenarios further increases the difficulty of understanding user intents\cite{ref85}. To address this, CPN can leverage ubiquitous computing resources to assist LAE nodes in generating communication policies, and utilize its programmable networking capabilities to dynamically configure instruction delivery strategies according to different scenarios, thereby improving the accuracy of LAE generative intelligence in understanding user intents. For example, the authors in \cite{ref80} proposed a cooperative AAV and edge devices based robust multi-access edge computing framework, leveraging generative AI-enhanced heterogeneous agent proximal policy optimization to jointly optimize the task offloading and AAV trajectory. The authors in \cite{ref81} proposed an AAV-edge collaborative multi-scenario adaptation framework for LAE generation intelligence, in which the generative models are decomposed into multiple customized sub-models, each specializing in the interpretation of service intents in different domains. Edge devices then select an appropriate sub-model for the AAV according to its operating scenario, ensuring that the AAV can output customized communication strategies.

In term of air-ground collaborative control, decision-making intelligence is mainly responsible for formulating communication strategies, AAV networking planning, and communication service planning. Considering the label-free learning nature of LAE decision intelligence tasks, DRL-based decision models are widely adopted\cite{ref86,ref87}. However, the limited resource and performance of LAE nodes often leads to high decision delay\cite{ref88}, while the lack of global AAV awareness makes it difficult to design reasonable AAV networking schemes\cite{ref89}. To address these challenges, on the one hand, CPN can migrate complex decision intelligence tasks to edge nodes or the cloud through computation offloading and distributed scheduling, thereby reducing AAV burden while ensuring low delay. On the other hand, CPN can aggregate AAV information collected by ubiquitous computing nodes in the cloud and assist AAV networking. For example, the authors in \cite{ref90} proposed a cloud–edge collaborative AAV control framework, in which AAV trajectory control tasks are offloaded to ground devices to reduce task execution latency. In this framework, edge devices are responsible for data preprocessing and caching, while the cloud handles the generation and dispatch of control commands. The authors in \cite{ref91} proposed an air-ground joint AAV networking framework, in which ground devices collect AAV information to formulate the network control problem and decompose it into a set of sub-problems. Distributed AAVs then solve individual sub-problems and generate corresponding control instructions.
\begin{figure*}[!t]
\centering
\includegraphics[width=\textwidth]{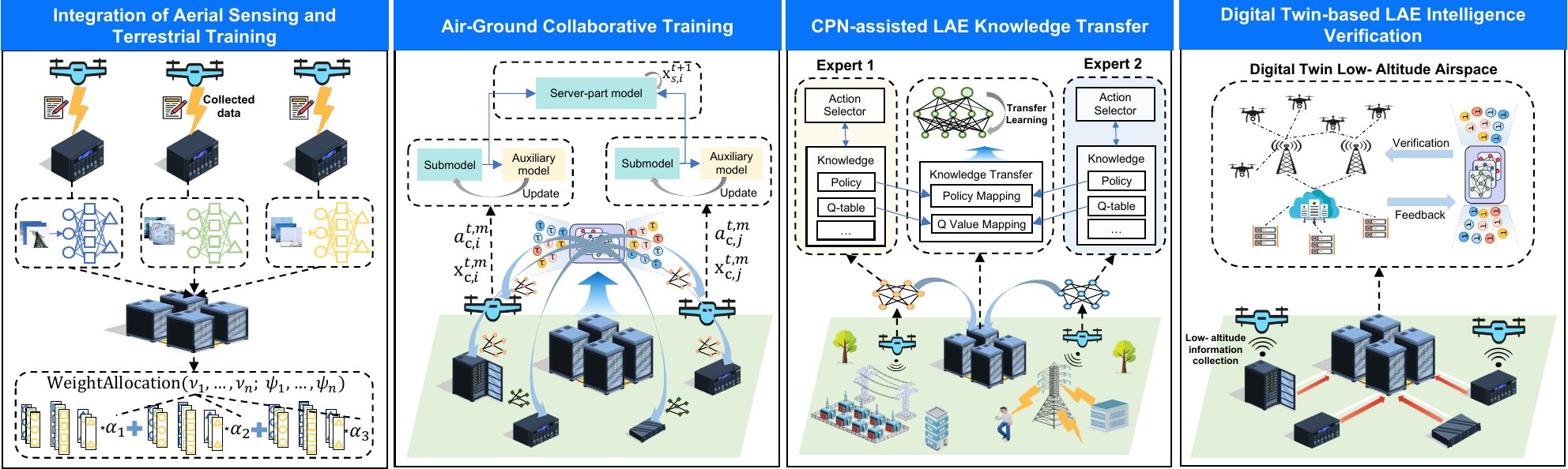}
\caption{CPN-assisted LAE intelligence training methods, including integration of aerial sensing and terrestrial training [82], air-ground collaborative training [85], CPN-assisted LAE knowledge transfer [50] and digital twin-based LAE intelligence verification [87]}
\end{figure*}
\subsection{CPN-assisted LAE Intelligence Training}
As shown in Fig. 2, CPN-assisted LAE intelligence training mainly involves classification and generation intelligence tasks in LAE communications. Due to the high dynamics of wireless channels and the complexity of interference factors in LAE networks, traditional rule-based sensing and classification methods struggle to adapt to rapid environmental changes. Classification intelligence can analyze received signal features in real time, enabling automatic classification of channel states, interference types, and other factors\cite{ref43,ref48}. However, considering the mobility of LAE nodes, the classification intelligence for LAE wireless sensing must cope with diverse scenarios. However, the number and scale of AI models that can be deployed on LAE nodes are limited, and these models are typically trained for specific scenarios. As a result, their sensing performance tends to degrade in unfamiliar environments. CPN can address this issue in two ways. First, it can leverage widely distributed edge and centralized computing resources to assist LAE nodes in training or inference of classification intelligence. For example, the authors in \cite{ref49} proposed a cloud-edge collaborative federated learning framework to tackle the multi-label classification problem in AAV broadband spectrum sensing. In this framework, computing resources at the end-side is used to collect data from various scenarios and train classification intelligence accordingly, while the central server determines model aggregation weights based on wireless channel features of different scenarios and aggregates model parameters from distributed sources. Second, CPN can make use of high-performance data centers to collect large volumes of LAE wireless sensing data and perform model transfer and generalization training for classification intelligence. For example, the authors in \cite{ref50} proposed a transfer learning framework for multi-AAV knowledge sharing, where AAVs act as distributed agents to collect sensing data across regions. A centralized server gathers their data and knowledge to conduct transfer learning, resulting in more general-purpose AI models.

Considering the complexity of training generative models, using LAE nodes for training is inefficient\cite{ref92}. In addition, generative models are prone to hallucination issues, and how to verify the validity of the trained models is also a problem that needs to be addressed\cite{ref93}. To tackle these challenges, CPN can, on the one hand, leverage high-performance computing power for large-scale generative model training or utilize distributed computing resources for personalized model training; on the other hand, it can build digital twin networks in data centers to validate the effectiveness of the trained generative models. For example, the authors in \cite{ref94} proposed an intelligent LAE service framework based on generative intelligence, which adopts a split learning strategy to distribute generative model modules and training tasks across edge devices, aerial devices, and the cloud, thereby fully leveraging heterogeneous computing resources, while also supporting user feedback-driven continuous model optimization. The authors in \cite{ref95} proposed an AAV-edge joint multi-agent generative AI training framework for cooperative task allocation, trajectory planning, and power management optimization in 6G-based air-ground integrated networks. The authors in \cite{ref96} proposed a digital twin simulation-enhanced generative AI training optimization framework, which leverages high-performance computing devices to construct a digital twin network that provides a safe and realistic training and testing environment for generative AI. At the same time, it enables the creation of diverse and adversarial scenarios for simulation-based robust agent training. The proposed framework leverages complex scenario generation capabilities, achieving a 15\% reduction in convergence time, approximately 21.4\% improvement in throughput under interference environments, and about 50\% increase in throughput in complex scenarios. It is foreseeable that this method can be effectively applied to LAE generative intelligence to provide a model validity testing environment. In Fig. 4, we summarize the above CPN-assisted LAE intelligent training methods and illustrate the operational flow of each method.

\subsection{Communication-Computing Collaborative Optimization}
As shown in Fig. 2, communication-computing collaborative optimization mainly involves generation and decision-making intelligence tasks in LAE communications. In this part, both intelligence serve the role of generating control decision for the system. In addition, generation intelligence can also perform certain communication network sensing tasks, such as radio map generation. By integrating them with CPN, communication-computing collaborative optimization can be achieved, thereby improving the QoS of low-altitude services. The main aspects include:

\subsubsection{Communication-aware Dynamic Resource Scheduling and Task Offloading} AAVs first perceive the low-altitude communication network state in real time and transmit it to the network control node. The network control node, considering factors such as bandwidth and interference in conjunction with computing load, decides whether the low-altitude service should be executed locally at the AAV node or offloaded to the edge or the cloud, while simultaneously determining the resource allocation scheme. For example, the authors in \cite{ref97} proposed a low-altitude MEC service execution optimization framework, where vertical collaboration between AAVs and edge devices, as well as horizontal collaboration among AAVs, is leveraged to improve service execution efficiency. In this network, the authors took into account the computing load of AAVs and edge devices, along with bandwidth and interference in the low-altitude communication network, to optimize AAV deployment and offloading decisions. They further propose a two-layer optimization algorithm, employing alternating optimization to address the non-convexity.

\subsubsection{Integrated Aerial Sensing, Communication and Computing} Integrated aerial sensing, communication and computing leverages AAV communication signals to perform sensing tasks, jointly considering sensing tasks and communication resources. Meanwhile, suitable computing power nodes are selected for processing according to the scale of the sensing data, thereby realizing a closed loop of sensing–communication–computing–control. For example, the authors in \cite{ref98} proposed a integrated AAV sensing, communication, and ground MEC framework, in which AAVs provide communication services for ground users while sensing multiple targets, with the sensing data offloaded to edge nodes for computation. The authors jointly optimize transmit/receive beamforming, AAV trajectories, and edge computing resources to enhance the efficiency of sensing data processing.

\subsubsection{Multi-AAV Collaborative Computing} AAV swarm can decompose complex tasks and perform distributed processing. However, the instability of inter-AAV communication links and the heterogeneity of AAV computing capabilities make multi-AAV task planning challenging. CPN, leveraging its programmable networking capabilities, can dynamically adjust multi-AAV collaboration schemes by jointly considering the state of aerial communication links and the computing loads of AAVs. For example, the authors in \cite{ref99} proposed a multi-agent optimization-based AAV cooperative planning method, in which AAVs are responsible for collecting surrounding environmental data, while multiple distributed agents together with a centralized critic jointly conduct AAV cooperative planning training and provide collaboration instructions for the AAVs.
\begin{figure}[!t]
\centering
\includegraphics[width=3in]{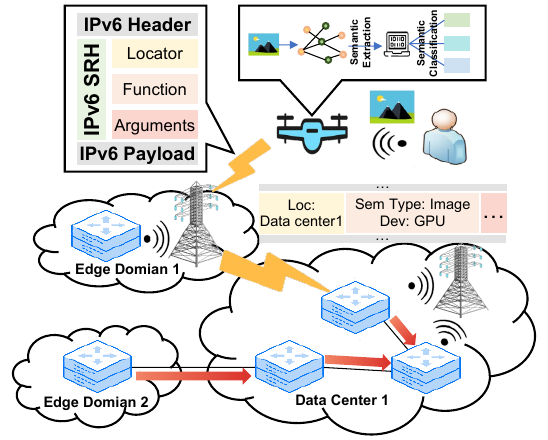}
\caption{The integration of LAE semantic classification and SRv6-based computing-aware routing. LAE nodes can encapsulate semantic types and computing power demands into computing-aware routing packets, enabling the efficient delivery of heterogeneous semantic data[94],[95].}
\end{figure}
\subsection{Low-Altitude Routing Optimization}
As shown in Fig. 2, low-altitude routing optimization mainly involves classification and prediction intelligence tasks in LAE communications. Semantic communication can significantly improve the efficiency of AAV image transmission \cite{ref52}. In LAE semantic transmission, classification intelligence can be employed to categorize different semantic information and transmit them to appropriate target nodes, thereby improving the accuracy of semantic reconstruction or the effectiveness of data analysis \cite{ref53,ref54}. However, in scenarios like large-scale monitoring, a significant challenge arises from the mixed transmission of heterogeneous semantic data, which varies in both type and processing requirements. Determining the appropriate nodes for semantic recovery and data analysis becomes a key issue faced by classification intelligence in LAE systems. To address this, the Computing-Aware Routing technology in the CPN offers a promising solution. Unlike traditional routing that only considers latency and reliability, CAR takes into account the computational demands of the transmitted data \cite{ref13}. Routing decisions are made not just based on network conditions, but also on the computing capabilities required by the semantic information. For example, the authors in \cite{ref56} proposed an SDN-based routing optimization framework for CPN, where a software-defined control plane gathers real-time information on computing resources and network connectivity. It formulates routing strategies by jointly considering service computing requirements and delay constraints. The authors in \cite{ref57} introduced a programmable CPN forwarding and scheduling method based on SRv6. This approach appends service type and computational demand information to the programmable segments of SRv6 packets. It then applies a user satisfaction function to match users with appropriate computing power nodes, while using the OSPF algorithm to determine optimal routing paths. The proposed routing method improves the forwarding and processing delay by approximately 43\%. Similarly, classification intelligence in LAE communication networks can encapsulate semantic types and computing power demands into SRv6-based computing-aware routing packets, as shown in Fig. 5. This enables the efficient delivery of heterogeneous semantic data to suitable processing nodes without requiring the LAE node to make complex decisions about destination selection. As a result, this helps reduce the operational burden on LAE nodes, allowing them to focus on classification tasks. 

Prediction intelligence can be applied to forecast users’ QoS requirements for low-altitude services, thereby enabling personalized service provision\cite{ref101}. However, in air–ground collaborative scenarios with large-scale task concurrency, the strong diversity of user demands raises the challenge of how to quickly match computing nodes based on prediction results and transmit data to the target node. To address this, CAR is clearly a promising solution, as it can match the required types and amounts of computing resources to services according to users’ QoS demands and enable efficient data transmission. For example, the authors in \cite{ref100} proposed a QoS-aware computing power routing framework, where a centralized controller collects service demand and network state information, and devises personalized computing resource allocation and routing schemes for services with different requirements. The framework also supports multiple QoS metrics, such as delay, jitter and availability, with adjustable weights, comprehensively considering various QoS dimensions to determine the optimal route for each service.

\subsection{Software-defined LAE Intelligent Communication}
As shown in Fig. 2, software-defined LAE intelligent communication mainly involves classification and decision-making intelligence tasks in LAE communications. Considering the wide coverage of LAE communication networks, LAE nodes often need to handle data transmissions of a large number of services with different types and demands\cite{ref5}. In response, it is necessary to consider using classification intelligence to classify services or traffic in LAE networks, so as to develop targeted data transmission policies\cite{ref58,ref59}. However, after service analysis, how LAE nodes can design customized communication policies or resource allocation schemes according to the demands of different services is also a problem that needs to be addressed. Considering the software-defined networking capability of CPN\cite{ref12}, LAE nodes can offload service analysis and classification results to the software-defined control plane in the CPN, which can customize communication policies and resource allocation schemes as needed, thereby enabling dynamic and customized management of LAE communication networks. For example, the authors in \cite{ref63} proposed a programmable aerial-ground wireless network, where airborne nodes form a mobile plane to collect service data, and ground computing nodes form a software-defined control plane to dynamically manage AAV communication networks based on customized network control and optimization parameters. The authors in \cite{ref102} proposed a software-defined integrated AAV and MEC framework, in which the control plane dynamically configures low-altitude service forwarding and resource allocation rules based on network state and multiple evaluation metrics. Furthermore, they proposed a DRL-SDNC algorithm that allocates bandwidth, computing, and storage resources according to task requirements, tolerable delay, and network conditions, thereby improving resource utilization and QoS in MEC-AAV environments. The proposed algorithm improves resource efficiency by 66.6\% and achieve better balance in multi-objective optimization.

As an important application of decision-making intelligence, intelligent service planning can dynamically design service chains and resource orchestration schemes\cite{ref103}. The SDN capability and computing-aware service orchestration capability of CPN can be combined with decision-making intelligence to provide flexible and controllable service optimization support for low-altitude communication tasks. The authors in \cite{ref104} proposed an AAV service orchestration framework in an MEC NFV scenario, where an integer linear programming model is designed for the workflow of the NFV orchestrator. The model aims to minimize the deployment cost of low-altitude services in edge cloud infrastructures.

\subsection{Ubiquitous Processing of Low-Altitude Information}
As shown in Fig. 2, the ubiquitous processing of low-altitude information involves all four kinds of intelligence. This is because in scenarios with a large number of concurrent services, the limited data processing capability of LAE results in high latency for information processing\cite{ref105,ref106}. Facing this challenge, CPN can leverage its widely distributed computing power nodes to assist LAE nodes in information processing. LAE nodes can also offload part of the tasks to distributed computing power nodes to reduce service analysis latency. For example, the authors in \cite{ref61} proposed an image service processing framework based on the collaboration between AAV and edge nodes. In this framework, the AAV offloads part of the image service analysis tasks to edge nodes, which complete the analysis and send control instructions back to the AAV. Similarly, the authors in \cite{ref62} proposed a task-oriented edge-aerial collaboration framework for efficient image service analysis. They also designed a feature pruning method based on an orthogonally-constrained variational information bottleneck encoder to reduce the transmission data volume between the AAV and edge nodes, thereby improving communication efficiency. The authors in \cite{ref107} proposed an edge computing-assisted AAV autonomous flight framework, in which AAVs offload visual data, communication capacity, and other information to edge nodes for 3D map generation and online trajectory planning.

In addition to leveraging computation offloading to assist LAE nodes in information processing, CPN can also support the execution of low-altitude services through its computing-network integration capability\cite{ref108}. By utilizing in-network computing technologies and computing-transmission integrated hardware such as DPU, CPN can perform data preprocessing while simultaneously ensuring fast transmission of low-altitude services. For example, the authors in \cite{ref109} proposed an AAV image data transmission preprocessing method, which selects the parts of the image that contribute most to the classifier under various channel conditions at the sender side, thereby reducing the amount of data transmitted and improving classification accuracy. The authors in \cite{ref110} proposed a real-time remote sensing data transmission preprocessing framework that leverages DPUs to perform data filtering, load balancing, and parallel processing during forwarding, thereby accelerating remote sensing data analysis tasks. Considering that LAE nodes can efficiently execute remote sensing tasks due to their high mobility, this method can foreseeably be integrated with LAE.
\subsection{Lessons Learned}
The distributed, programmable, and computing-aware nature of CPNs is essential for overcoming the inherent limitations of lightweight aerial nodes. By leveraging the global resource orchestration of CPNs, LAE networks can achieve more accurate control decisions, scalable intelligence training, and joint optimization of communication and computation. This suggests that future LAE systems should not only rely on localized aerial processing but also actively exploit the ubiquitous computing and service orchestration capabilities of CPNs to ensure low-latency, reliable, and adaptive communication in dynamic environments.

\begin{table*}[!t]
\centering
\caption{Summary of Related Works in LAE Intelligent Communications for CPN}
\renewcommand{\arrayrulewidth}{0.8pt}
\renewcommand{\tabcolsep}{10pt}
{\fontsize{8}{10}\selectfont
\begin{tabular}{m{2cm}|m{2cm}|m{12cm}}
    \hline
    \rowcolor{gray!20}
    \multicolumn{1}{c|}{\textbf{Topic}} & 
    \multicolumn{1}{c|}{\textbf{References}} & 
    \multicolumn{1}{c}{\parbox{12cm}{\textbf{Challenges of the CPN, techniques and advatages of proposed methods}}} \\ 
    \hline
    Air-ground collaborative control & [115], [116], [119], [120] & \makecell[l]{$\times$ \textbf{Challenge}: The adaptability of CPN ground nodes is insufficient, and fixed deployment leads to slow\\ response to hotspots, making resource fragmentation difficult to coordinate across domains. \\ \checkmark \, \textbf{Techniques}: Aerial mobile coverage, real-time sensing, distributed coordination. \\ \checkmark \, \textbf{Advantage}:  Realizing dynamic balance of cross domain resources, enhance the f\textbf{}lexibility and real-time\\ performance of global control.}  
     \\ 
    \hline
    Air-ground collaborative LAE intelligence training & [123], [124], [126], [127] & \makecell[l]{$\times$ \textbf{Challenge}: CPN models based on ground data lack generalization ability for dynamic airspace, and \\centralized training incurs high costs and delays. \\ \checkmark \, \textbf{Techniques}: \, Aerial perception data collection, distributed learning, model splitting and transfer. \\ \checkmark \, \textbf{Advantage}: Improving the adaptability of the training model to dynamic environments, accelerate\\ convergence speed, and reduce communication burden.}
     \\ 
    \hline
    Communication-computing collaborative optimization & [129], [130], [132], [133] & \makecell[l]{$\times$ \textbf{Challenge}: The separation of communication and computation scheduling in CPN makes it difficult to\\ quickly match dynamic hotspots, which can lead to congestion and imbalanced computing power. \\ \checkmark \, \textbf{Techniques}: \, Real time network state perception, air ground collaborative load balancing, multi-AAV \\collaborative scheduling. \\ \checkmark \, \textbf{Advantage}: Strengthening communication computing collaboration, improve system resource utilization\\ and QoS assurance.} \\
    \hline
    Low-altitude routing optimization & [135], [136], [138], [139] & \makecell[l]{$\times$ \textbf{Challenge}: CPN ground routing equipment is fixed, making it difficult to cope with coverage blind\\ spots and sudden business hotspots, and traditional routing is difficult to adjust quickly. \\ \checkmark \, \textbf{Techniques}: \, Dynamic deployment, aerial relay, real-time link sensing. \\ \checkmark \, \textbf{Advantage}: Enhancing routing flexibility and fault tolerance, reduce latency\textbf{}, and improve reliability.}\\
    \hline
    Software defined LAE intelligent communication & [141]-[144] & \makecell[l]{$\times$ \textbf{Challenge}: CPN lacks cross layer coordination and separates communication, computation, and service\\ orchestration, resulting in overall performance limitations. \\ \checkmark \, \textbf{Techniques}: \, Network function virtualization, dynamic programming network, cross-layer perception. \\ \checkmark \, \textbf{Advantage}: Realizing cross layer optimization and business aware network management, enabling\\ CPN to have stronger adaptability and programmability.}\\
    \hline
\end{tabular}}
\end{table*}

\section{LAE Intelligent Communications for CPN}
Conversely, LAE intelligent communications can also provide flexible auxiliary support for CPN. Leveraging its wide aerial coverage and high mobility, LAE can offer functionalities such as computational power supplementation, data relaying, and service collaboration, thereby enhancing the resource scheduling and service assurance capabilities of computing power networks in dynamic environments. In this section, we introduce how LAE supports the new capabilities in Fig. 2 and assists the operation of CPN. We summarize the related works in Table VI.
\subsection{Air-Ground Collaborative Control}
One major challenge in CPN control arises from the limited adaptability of ground computing devices when facing highly dynamic service demands\cite{ref136}. The resource allocation and control decisions of CPN rely on ground nodes and centralized coordination. However, due to fixed deployment locations and static deployment density, computing nodes are often insufficient to provide timely responses when service hotspots, e.g., in a remote area due to disaster/accident or special events, emerge or shift\cite{ref137}. As a result, service responses and control signal transmissions may encounter higher latency, reducing the responsiveness of global control. This issue is particularly pronounced in scenarios such as mobile edge computing or emergency services requiring rapid task offloading and rerouting. LAE intelligent communications can directly address these weaknesses by serving as a mobile extension of the CPN control plane. With wide-area aerial coverage and adaptive repositioning capabilities, LAE nodes can fill coverage gaps, collect up-to-date network states, and forward control signals through low-latency aerial links. For example, the authors in \cite{ref124} proposed an AAV-assisted joint resource management and service offloading framework for computing services in remote areas, where they also introduced a collaborative computation offloading scheme that considers interference suppression between AAVs and edge computing devices. The authors in \cite{ref125} proposed an AAV–edge computing collaborative framework for dynamic environment information collection and processing, in which AAVs are responsible for sensing large-scale environmental information, while edge computing devices receive these data for processing and control decision-making. Meanwhile, the authors introduced an AAV control algorithm designed to maximize the efficiency of AAV data collection and processing under constraints such as real-time data processing, obstacle avoidance, and system mobility. The proposed method reduced energy consumption by about 44\% and increased data acquisition by about 17\%.

Another challenge for CPN control stems from resource heterogeneity and fragmentation, which complicates cross-domain control and collaboration\cite{ref138}. CPN integrates multi-layer computing and networking resources, including end devices, edge servers, and cloud facilities. However, the lack of flexible coordination entities may lead to resource silos, resulting in load imbalance and underutilization of capacity. Since centralized controllers struggle to scale under massive tasks and diverse QoS requirements, relying solely on ground infrastructure is insufficient for real-time and effective coordination of these heterogeneous resources\cite{ref139}. LAE intelligent communication nodes can help mitigate these issues by acting as mobile distributed coordinators. These coordinators can interconnect isolated resource clusters and enhance control dynamism. Equipped with intelligent communication functions, aerial nodes can adaptively establish temporary links between underutilized and overloaded areas, balance workloads, and disseminate synchronization information across domains. For example, the authors in \cite{ref126} proposed a multi-AAV-enabled collaborative edge computing framework, where AAVs dynamically coordinate heterogeneous edge computing resources and construct air–ground and ground–ground collaborative service links. They further introduced a two-layer optimization framework to optimize AAV deployment and resource allocation. The authors in \cite{ref127} proposed a multi-AAV-assisted large-scale MEC framework, which jointly optimizes the associations between AAVs and ground edge computing devices, the cooperation of heterogeneous computing devices, and AAV deployment, with the goal of minimizing overall energy consumption.

\subsection{Air-Ground Collaborative LAE Intelligence Training}
One challenge in CPN-based AI training is the limited generalization capability of models trained solely on ground-based data and computing infrastructure. Although CPN integrates large-scale computing and networking resources, most training tasks are concentrated in static data centers or regional edge servers\cite{ref140}. These environments cannot capture dynamic airspace conditions, such as channel fluctuations caused by mobility, topology changes, and the diverse service demands. Moreover, centralized training often lacks real-time feedback from the rapidly evolving air–ground environment, restricting its ability to adapt continuously\cite{ref141}. LAE intelligent communications can overcome these limitations by actively involving aerial nodes in the distributed training process. With onboard sensing and computing capabilities, LAE nodes can generate real-time data samples reflecting unique airspace characteristics and transmit them to the CPN training framework. For instance, the authors in \cite{ref128} proposed an AAV surveillance framework integrated with sub-agent models, where AAVs deploy sub-agent models from a ground central controller to monitor, search, and track specific areas, while customizing the training of these models for different task types assigned by the central controller. The authors in \cite{ref129} introduced an AAV-assisted self-supervised learning method for ground computing devices to optimize terrain perception in off-road navigation. AAVs collect aerial terrain images, which are combined with ground vehicle vibration and energy consumption data to train models predicting terrain characteristics.

Another challenge lies in the inefficiency and scalability bottlenecks of centralized AI training within CPN. Training large-scale models in centralized data centers requires aggregating massive amounts of distributed data, which incurs significant communication overhead, high latency, and potential privacy risks\cite{ref142}. These limitations slow down model convergence and reduce the practicality of deploying intelligent training for time-sensitive LAE applications. LAE intelligent communications address this issue by providing a collaborative training layer that dynamically balances workloads between aerial and ground nodes. LAE nodes can offload part of the training computations, relay intermediate model updates, and even perform preliminary training tasks before aggregation in the CPN. This significantly reduces communication burdens while accelerating model convergence. For example, the authors in \cite{ref130} proposed an AAV-assisted event recognition model-splitting federated learning framework, where computational workloads are distributed between AAVs and the central server, optimizing device efficiency and global model performance. Lightweight convolutional models extract spatiotemporal features from video sequences captured by AAVs, reducing computational complexity and transmission overhead. The proposed method reduced memory usage by an average of 74.18\% and computation latency by 74.89\%. The authors in \cite{ref131} introduced an air–ground collaborative federated learning scheme based on device-to-device communication to address heterogeneity in AAV-assisted AI training, allowing edge devices and AAVs to migrate non-private data samples between each other for more efficient AI training.

\subsection{Communication-Computing Collaborative Optimization}
A key challenge in communication–computing collaborative optimization within CPN is the mismatch between resource availability and dynamic service hotspots\cite{ref143}. In large-scale networks, computing resources and communication links are often provisioned based on historical traffic patterns or static planning, which cannot quickly adapt to sudden surges in task loads or shifts in user locations. This mismatch may cause node overloading or resource underutilization, thereby reducing overall system efficiency. Furthermore, network congestion and variable latency in ground infrastructure can delay the feedback required for joint optimization of computation and communication, undermining QoS guarantees. With flexible deployment and wide-area coverage, LAE nodes can monitor real-time network conditions, relay critical state information to CPN controllers, and even perform localized optimization for load balancing and task offloading. For instance, the authors in \cite{ref132} proposed an AAV-assisted ground facility and wireless network state sensing framework, integrating AAV path planning, dynamic data collection, and node scheduling mechanisms to support network management for ground devices. The proposed framework can sense state information such as communication link state, length of messages and queue. The authors in \cite{ref133} presented an AAV-assisted computing resource management framework, where AAVs act as network information collectors and apply a weighted-sum method to optimize computing device selection, AAV relay assignment, and source power allocation.

In addition, the weak coupling between communication and computing resource management in CPN poses another challenge for communication–computing collaborative control. In many current CPN architectures, communication scheduling, such as bandwidth allocation and routing, and computing scheduling, such as task offloading and workload balancing, are handled independently, which hinders the system’s ability to achieve joint optimization of resources across heterogeneous domains\cite{ref144}. To address this, LAE nodes can jointly sense link conditions and computational loads, enabling more comprehensive optimization. For example, the authors in \cite{ref134} proposed an AAV-assisted mobile crowdsensing framework, where AAVs are responsible for sensing the wireless network environment and collecting data. The framework jointly considers device computing capabilities and a directional energy transmission model to optimize resource allocation strategies and AAV data collection volume. The authors in \cite{ref135} presented an AAV–MEC optimization scheme integrated with RIS, where AAV equipped with RIS constructs multidimensional signal paths to suppress eavesdropping channels while enhancing legitimate link gains, and simultaneously optimizes ground user transmit power, AAV trajectories, and edge computing resource allocation.

\begin{figure}[!t]
\centering
\includegraphics[width=\columnwidth]{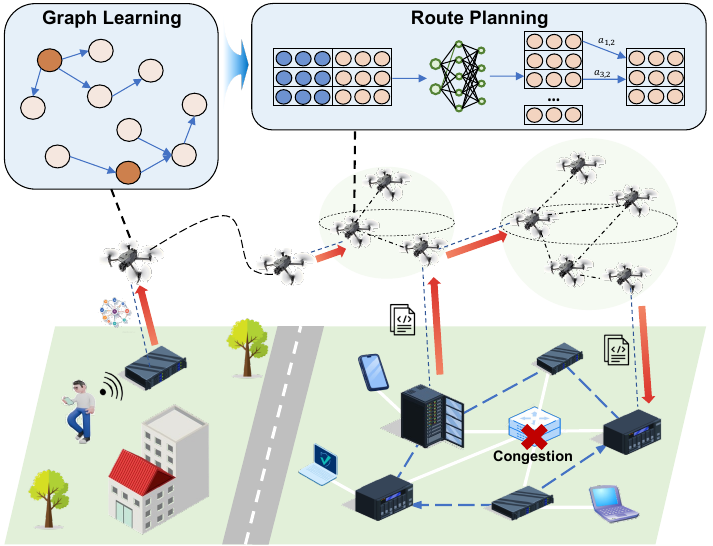}
\caption{Low-altitude routing optimization. With mobility and altitude advantages, AAVs can be efficiently deployed to supplement routing nodes, build temporary communication networks, and establish optimized air–ground routing paths to improve the CPN routing [134], [135].}
\end{figure}

\subsection{Low-Altitude Routing Optimization}
A key challenge in CPN routing is the lack of flexibility in ground routing devices when facing coverage gaps or surging computational service demands\cite{ref149}. Ground routing relies on fixed base stations and backbone nodes, whose deployment density and geographic distribution cannot promptly respond to shifting service hotspots. This results in coverage blind spots, imbalanced traffic loads, and inefficient routing paths, ultimately degrading computational service quality. To overcome this limitation, LAE intelligent communication nodes can serve as aerial base stations, providing flexible coverage extension and dynamic aerial routing support, as shown in Fig. 6. With mobility and altitude advantages, AAVs can be efficiently deployed to supplement routing nodes, build temporary communication networks, and establish optimized air–ground routing paths to reduce latency. For example, the authors in \cite{ref145} proposed a graph reinforcement learning based multi-hotspot region AAV dynamic scheduling framework, as shown in Fig. 6, where AAVs learn traffic patterns and trends and dynamically fly to hotspot regions to provide wireless connectivity. They abstracted the dynamic scheduling among multiple hotspots into a dynamic topology graph and leveraged graph convolution to capture spatial–temporal relations between hotspots. In \cite{ref146}, the authors presented a predictive AAV base station deployment and service offloading framework, where AAVs forecast ground data traffic and its impact on computational facilities, and then dynamically act as aerial base stations to provide connectivity and service, alleviating high-demand or poor-connectivity regions. Compared with ground routing, this approach achieved a 31\% throughput improvement and enhanced network service level.

Another challenge in CPN routing arises from its static and centralized decision-making process, which often fails to adapt to traffic fluctuations and unpredictable service demands\cite{ref150}. Traditional routing schemes usually optimize paths based on relatively stable network states, making them inefficient in the presence of real-time variations such as node failures, interference, or sudden workload surges. This can lead to congestion and latency, which are intolerable for delay-sensitive computational services. To address this, AAV-assisted low-altitude routing offers a promising solution. On the one hand, AAVs can perform real-time sensing to detect local network congestion or link failures and dynamically adjust routing paths. On the other hand, AAVs can act as mobile relays, rerouting CPN services to appropriate nodes and thereby enhancing reliability. For example, the authors in \cite{ref147} proposed an AAV-assisted ground facility routing protocol, where AAVs dynamically monitor the status of ground communication links and serve as relays to improve routing and connectivity under congestion or poor network conditions. The authors in \cite{ref148} studied AAV-assisted access networks based on link path loss, where AAVs sense channel indicators such as shadow fading and leverage improved AAV-Dijkstra algorithm to optimize routing between ground devices by incorporating path loss into the decision-making process.

\subsection{Software Defined LAE Intelligent Communication}
A key challenge for CPN management lies in the lack of cross-layer coordination, where communication, computation, and service orchestration are managed independently, leading to suboptimal overall performance\cite{ref155}. This separation makes it difficult for CPNs to jointly optimize transmission efficiency and computational service orchestration, especially under delay-sensitive and resource-constrained conditions. Software-defined LAE intelligent communication provides a solution by enabling fine-grained cross-layer programmability. Specifically, LAE nodes equipped with software-defined modules can dynamically sense computational workloads, network congestion, and service requirements, and then reprogram routing, resource allocation, and computation offloading strategies accordingly\cite{ref156}. This cross-layer programmability transforms the CPN from a rigid, device-centric model into a service-aware, intent-driven system. For example, the authors in \cite{ref151} integrated software-defined communication hardware into AAVs to perform real-time spectrum allocation for ground users, alleviating the computational load on ground control stations. This approach combines dynamic spectrum management with NFV to extend communication coverage and enhance quality. The authors in \cite{ref152} proposed leveraging software-defined air–ground integrated networks with in-network caching and computing to improve users’ video experience. They formulated the problem as joint optimization of video resolution adaptation, computation scheduling, and bandwidth provisioning, solved with the alternating direction method of multipliers. The authors in \cite{ref153} proposed a software-defined air–ground integrated network architecture, where both aerial and terrestrial resources were sliced to achieve service isolation, and pooled into a hierarchical resource management framework for ground services. The authors in \cite{ref154} presented an on-demand computation offloading strategy in AAV-aided MEC systems, where AAVs dynamically infer user intent and predict mobility, adapting computation offloading strategies in real time. Using a block coordinate descent method, the approach reduced path loss by 18.55\% and total energy consumption by 18.28\%.
\subsection{Lessons Learned}
The mobility, aerial coverage, and real-time sensing capabilities of LAE intelligent communications can compensate for the limitations of the static deployment of ground-based CPN infrastructures. By serving as a mobile extension of the control plane, enabling aerial data collection, and facilitating distributed collaboration, LAE can greatly enhance the adaptability of CPNs to dynamic hotspots and emergency situations. This indicates that future CPN designs should integrate LAE intelligent communications to form flexible and collaborative architectures, thereby achieving resilient, adaptive, and service-aware computing power networks.
\section{Future Directions}
In this section, we present important future directions for the synergy of LAE intelligent communication and CPN, which aim to tackle the joint challenges in Fig. 3 by advancing cloud–edge–air coordination, enhancing reliability and security mechanisms, and promoting energy-efficient designs to ensure resilient LAE–CPN integration.
\subsection{Digital Twin-enhanced LAE-CPN Integrated System}
The LAE-CPN integrated system lacks effective tools to predict network dynamics and validate new control policies before real deployment, which often results in inefficient routing or orchestration decisions under rapidly changing scenarios\cite{ref157}. Digital twin technology offers a promising solution by creating a virtual replica of the integrated LAE and CPN system, capable of simulating mobility patterns, traffic fluctuations, and service requests in real time. By synchronizing with real-world data, the digital twin enables proactive stress testing and scenario analysis, allowing operators to optimize resource allocation and control strategies before applying them in practice. This approach not only reduces risks but also accelerates the deployment of new algorithms and policies in highly dynamic environments\cite{ref158}.

\subsection{Security and Privacy in LAE-CPN Integrated System}
Although there are currently some CPN security and privacy protection methods designed for stable terrestrial infrastructures, they are not suited for the highly dynamic, distributed, and mobile nature of LAE nodes\cite{ref160}. This gap exposes the system to vulnerabilities such as spoofing of AAV control signals, eavesdropping on sensitive data offloaded from aerial devices, and privacy leakage during collaborative intelligence training. Future research should focus on developing lightweight yet robust security mechanisms tailored for LAE-CPN integration. Techniques such as blockchain-enabled trust management can provide decentralized authentication and verifiable transaction records among aerial and ground nodes, reducing reliance on vulnerable central controllers\cite{ref161}. Privacy-preserving federated learning frameworks can be applied to ensure that AAVs and edge nodes share only encrypted model parameters rather than raw data, mitigating risks of privacy leakage\cite{ref159}. Additionally, anomaly detection powered by AI can be integrated into AAVs to identify adversarial behaviors in real time, such as abnormal trajectory control or malicious traffic injection.

\subsection{Energy-aware LAE-CPN Integrated System}
A major problem in LAE-CPN systems is the mismatch between AAVs’ limited energy supply and the high energy demand of computation-intensive CPN tasks\cite{ref162}. Traditional resource allocation schemes largely ignore AAV energy states, which can lead to premature node failure and service disruption. Energy-aware resource allocation introduces mechanisms that jointly consider communication, computation, and AAV battery levels, potentially enhanced with energy harvesting and adaptive duty cycling. By dynamically offloading tasks between aerial and ground resources based on energy availability, the system can extend AAV operation time while still fulfilling computational requirements. This approach ensures sustainability and reliability of CPN services in energy-constrained low-altitude environments.
\subsection{Heterogeneous LAE nodes and Multi-Agent Coordination}
Although some existing solutions consider node heterogeneity in CPN, they still lack discussion on the heterogeneity of LAE nodes, this limits scalability and ignores practical differences in AAV hardware, sensing, and computing capabilities\cite{ref163}. In reality, the integrated LAE and CPN systems will involve heterogeneous AAV swarms ranging from lightweight drones to large high-end aerial platforms. Multi-agent coordination frameworks can address this issue by enabling AAVs to dynamically form coalitions, negotiate task distribution, and collaboratively support CPN services. Distributed learning and consensus-based decision mechanisms ensure that heterogeneous AAVs can cooperate efficiently without relying on centralized control. This approach improves scalability, fairness, and resilience, allowing the integrated LAE and CPN systems to support complex missions across diverse environments.
\subsection{Space-Air-Ground Integrated System}
A significant challenge in current LAE-CPN integrated systems is the limited global coverage and dependency on terrestrial infrastructures, which restricts connectivity in remote or disaster-affected regions. Ground and low-altitude nodes alone cannot guarantee seamless service continuity over wide geographic areas, especially when terrestrial networks are congested\cite{ref164}. Future research should explore the integration of space-based nodes, such as Low Earth Orbit satellites (LEO) \cite{ref165} or High-Altitude Platform Stations (HAPS)\cite{ref166}, with LAE and ground CPN infrastructures to form a three-tier Space-Air-Ground (SAG) system. In this architecture, satellites provide wide-area backbone connectivity and long-range data relay, AAVs offer flexible coverage and low-latency service augmentation, and ground nodes deliver high-capacity computing and storage. Joint optimization frameworks can coordinate resource allocation, routing, and service placement across these three layers. For example, satellite links could be dynamically leveraged for latency-tolerant data aggregation or model synchronization, while AAVs handle time-sensitive computations and relaying, and ground nodes execute intensive processing tasks.

\section{Conclusion}
In this paper, we have conducted a comprehensive analysis of the synergy between CPNs and LAE intelligent communications. First, we have introduced CPN and LAE intelligent communications, and analyzed the unique capabilities they offer compared with previous similar concepts. Based on these unique capabilities, we have further explored the new capabilities emerging from the synergy between CPNs and LAE intelligent communications, including air–ground collaborative control, CPN-assisted LAE intelligence training, communication–computing collaborative optimization, low-altitude routing optimization and software-defined LAE intelligent communication. Then We have analyzed the new challenges arising from the synergy of CPNs and LAE intelligent communications. Subsequently, we have discussed how CPNs and LAE intelligent communications support each other within the above new capabilities and reviewed relevant related work. Finally, we have outlined future research directions, highlighting the potential of digital twins, security and privacy protection, energy management, multi-agent systems and integrated space-air–ground technologies in CPN–LAE integrated systems.

\bibliography{reference}

\begin{thebibliography}{100}
\providecommand{\url}[1]{#1}
\csname url@samestyle\endcsname
\providecommand{\newblock}{\relax}
\providecommand{\bibinfo}[2]{#2}
\providecommand{\BIBentrySTDinterwordspacing}{\spaceskip=0pt\relax}
\providecommand{\BIBentryALTinterwordstretchfactor}{4}
\providecommand{\BIBentryALTinterwordspacing}{\spaceskip=\fontdimen2\font plus
\BIBentryALTinterwordstretchfactor\fontdimen3\font minus \fontdimen4\font\relax}
\providecommand{\BIBforeignlanguage}[2]{{%
\expandafter\ifx\csname l@#1\endcsname\relax
\typeout{** WARNING: IEEEtran.bst: No hyphenation pattern has been}%
\typeout{** loaded for the language `#1'. Using the pattern for}%
\typeout{** the default language instead.}%
\else
\language=\csname l@#1\endcsname
\fi
#2}}
\providecommand{\BIBdecl}{\relax}
\BIBdecl

\bibitem{ref5}
\BIBentryALTinterwordspacing
Y.~Wang, G.~Sun, Z.~Sun, J.~Wang, J.~Li, C.~Zhao, J.~Wu, S.~Liang, M.~Yin, P.~Wang, D.~Niyato, S.~Sun, and D.~I. Kim, ``Toward realization of low-altitude economy networks: Core architecture, integrated technologies, and future directions,'' 2025. [Online]. Available: \url{https://arxiv.org/abs/2504.21583}
\BIBentrySTDinterwordspacing

\bibitem{ref111}
H.~Huang, J.~Su, and F.-Y. Wang, ``The potential of low-altitude airspace: The future of urban air transportation,'' \emph{IEEE Transactions on Intelligent Vehicles}, vol.~9, no.~8, pp. 5250--5254, 2024.

\bibitem{ref20}
X.~Ye, Y.~Mao, X.~Yu, S.~Sun, L.~Fu, and J.~Xu, ``Integrated sensing and communications for low-altitude economy: A deep reinforcement learning approach,'' \emph{IEEE Transactions on Wireless Communications}, pp. 1--1, 2025.

\bibitem{ref6}
S.~Fu, M.~Zhang, M.~Liu, C.~Chen, and F.~R. Yu, ``Toward energy-efficient uav-assisted wireless networks using an artificial intelligence approach,'' \emph{IEEE Wireless Communications}, vol.~29, no.~5, pp. 77--83, 2022.

\bibitem{ref7}
Y.~Wang, Y.~He, F.~R. Yu, B.~Song, and V.~C. Leung, ``Efficient resource allocation for building the metaverse with uavs: A quantum collective reinforcement learning approach,'' \emph{IEEE Wireless Communications}, vol.~30, no.~5, pp. 152--159, 2023.

\bibitem{ref35}
\BIBentryALTinterwordspacing
G.~Luo, J.~Li, Q.~Zhang, Z.~Feng, Q.~Yuan, Y.~Lin, H.~Zhang, N.~Cheng, and P.~Zhang, ``Toward low-altitude airspace management and uav operations: Requirements, architecture and enabling technologies,'' 2025. [Online]. Available: \url{https://arxiv.org/abs/2506.08579}
\BIBentrySTDinterwordspacing

\bibitem{ref36}
P.~Cao, L.~Lei, S.~Cai, G.~Shen, X.~Liu, X.~Wang, L.~Zhang, L.~Zhou, and M.~Guizani, ``Computational intelligence algorithms for uav swarm networking and collaboration: A comprehensive survey and future directions,'' \emph{IEEE Communications Surveys \& Tutorials}, vol.~26, no.~4, pp. 2684--2728, 2024.

\bibitem{ref37}
\BIBentryALTinterwordspacing
X.~Gao, Y.~Wang, B.~Liu, X.~Zhou, R.~Zhang, J.~Wang, D.~Niyato, D.~I. Kim, A.~Jamalipour, C.~Yuen, J.~An, and K.~Yang, ``Agentic satellite-augmented low-altitude economy and terrestrial networks: A survey on generative approaches,'' 2025. [Online]. Available: \url{https://arxiv.org/abs/2507.14633}
\BIBentrySTDinterwordspacing

\bibitem{ref38}
Y.~Bai, H.~Zhao, X.~Zhang, Z.~Chang, R.~Jäntti, and K.~Yang, ``Toward autonomous multi-uav wireless network: A survey of reinforcement learning-based approaches,'' \emph{IEEE Communications Surveys \& Tutorials}, vol.~25, no.~4, pp. 3038--3067, 2023.

\bibitem{ref39}
\BIBentryALTinterwordspacing
Z.~Xu, T.~Zheng, and L.~Dai, ``Empowering near-field communications in low-altitude economy with llm: Fundamentals, potentials, solutions, and future directions,'' 2025. [Online]. Available: \url{https://arxiv.org/abs/2506.17067}
\BIBentrySTDinterwordspacing

\bibitem{ref40}
\BIBentryALTinterwordspacing
M.~Ahmed, A.~A. Soofi, F.~Khan, S.~Raza, W.~U. Khan, L.~Su, F.~Xu, and Z.~Han, ``Toward a sustainable low-altitude economy: A survey of energy-efficient ris-uav networks,'' 2025. [Online]. Available: \url{https://arxiv.org/abs/2504.02162}
\BIBentrySTDinterwordspacing

\bibitem{ref4}
X.~Gong, C.~Bai, S.~Ren, J.~Wang, and C.~Wang, ``A survey of compute first networking,'' in \emph{2023 IEEE 23rd International Conference on Communication Technology (ICCT)}, 2023, pp. 688--695.

\bibitem{ref12}
S.~Yukun, L.~Bo, L.~Junlin, H.~Haonan, Z.~Xing, P.~Jing, and W.~Wenbo, ``Computing power network: A survey,'' \emph{China Communications}, vol.~21, no.~9, pp. 109--145, 2024.

\bibitem{ref13}
X.~Tang, C.~Cao, Y.~Wang, S.~Zhang, Y.~Liu, M.~Li, and T.~He, ``Computing power network: The architecture of convergence of computing and networking towards 6g requirement,'' \emph{China Communications}, vol.~18, no.~2, pp. 175--185, 2021.

\bibitem{ref8}
H.~Chen, D.~Wang, J.~Fang, Y.~Li, S.~Xu, and Z.~Xu, ``Lw-yolov8: An lightweight object detection algorithm for uav aerial imagery,'' in \emph{2024 6th International Conference on Natural Language Processing (ICNLP)}, 2024, pp. 446--450.

\bibitem{ref9}
N.~Gao, Y.~Zeng, J.~Wang, D.~Wu, C.~Zhang, Q.~Song, J.~Qian, and S.~Jin, ``Energy model for uav communications: Experimental validation and model generalization,'' \emph{China Communications}, vol.~18, no.~7, pp. 253--264, 2021.

\bibitem{ref10}
Y.~Xi, J.~Chen, and X.~Tao, ``Obstacle detection and avoidance algorithm in uav patrol inspection of transmission lines based on multi-source data fusion and computer vision,'' in \emph{2024 International Conference on Telecommunications and Power Electronics (TELEPE)}, 2024, pp. 614--619.

\bibitem{ref11}
L.~Li, R.~Yang, M.~Lv, A.~Wu, and Z.~Zhao, ``From behavior to natural language: Generative approach for unmanned aerial vehicle intent recognition,'' \emph{IEEE Transactions on Artificial Intelligence}, vol.~5, no.~12, pp. 6196--6209, 2024.

\bibitem{ref14}
Y.~Liu, B.~Guo, N.~Li, Y.~Ding, Z.~Zhang, and Z.~Yu, ``Crowdtransfer: Enabling crowd knowledge transfer in aiot community,'' \emph{IEEE Communications Surveys \& Tutorials}, vol.~27, no.~2, pp. 1191--1237, 2025.

\bibitem{ref15}
Y.~Zhang, C.~Cao, X.~Tang, R.~Pang, S.~Wang, and X.~Wen, ``Programmable service system based on sidaas in computing power network,'' in \emph{2022 5th International Conference on Hot Information-Centric Networking (HotICN)}, 2022, pp. 67--71.

\bibitem{ref16}
Z.~Ming, Q.~Guo, H.~Yu, and T.~Taleb, ``Deep reinforcement learning-based task offloading over in-network computing and multi-access edge computing,'' in \emph{2023 International Conference on Networking and Network Applications (NaNA)}, 2023, pp. 281--286.

\bibitem{ref17}
Z.~Yang, S.~Bi, and Y.-J.~A. Zhang, ``Dynamic offloading and trajectory control for uav-enabled mobile edge computing system with energy harvesting devices,'' \emph{IEEE Transactions on Wireless Communications}, vol.~21, no.~12, pp. 10\,515--10\,528, 2022.

\bibitem{ref18}
S.~Jiang, Z.~Huang, and Y.~Ji, ``Adaptive uav-assisted geographic routing with q-learning in vanet,'' \emph{IEEE Communications Letters}, vol.~25, no.~4, pp. 1358--1362, 2021.

\bibitem{ref117}
G.~K. Pandey, D.~S. Gurjar, S.~Yadav, Y.~Jiang, and C.~Yuen, ``Uav-assisted communications with rf energy harvesting: A comprehensive survey,'' \emph{IEEE Communications Surveys \& Tutorials}, vol.~27, no.~2, pp. 782--838, 2025.

\bibitem{ref113}
Y.~Wang, J.~Huang, F.~Shan, Y.~Gao, R.~Xiong, and J.~Luo, ``Optimizing joint speed and altitude schedule for uav data collection in low-altitude airspace,'' \emph{IEEE Transactions on Mobile Computing}, pp. 1--15, 2025.

\bibitem{ref112}
S.~Zhao, L.~Wen, D.~Tang, T.~Wu, M.~Elkashlan, F.~Adachi, N.~Al-Dhahir, and C.~Yuen, ``Exploit security for low-altitude economy: A swipt-driven strategy with uav-mounted mf-ris,'' \emph{IEEE Transactions on Vehicular Technology}, pp. 1--6, 2025.

\bibitem{ref114}
Z.~Xu, T.~Zheng, and L.~Dai, ``Llm-empowered near-field communications for low-altitude economy,'' \emph{IEEE Transactions on Communications}, pp. 1--1, 2025.

\bibitem{ref21}
D.~He, W.~Yuan, J.~Wu, and R.~Liu, ``Ubiquitous uav communication enabled low-altitude economy: Applications, techniques, and 3gpp’s efforts,'' \emph{IEEE Network}, pp. 1--1, 2025.

\bibitem{ref1}
Z.~Zhang, J.~Wang, J.~Chen, Z.~Fang, C.~Jiang, and Z.~Han, ``A priority-aware ai-generated content resource allocation method for multi-uav aided metaverse,'' in \emph{2025 IEEE Wireless Communications and Networking Conference (WCNC)}, 2025, pp. 1--6.

\bibitem{ref19}
N.~Zhang, ``Research on lightweight ai models for uav identity authentication and intrusion detection in the low-altitude economy,'' in \emph{2025 International Conference on Artificial Intelligence and Engineering Management (ICAIEM)}, 2025, pp. 91--95.

\bibitem{ref22}
H.-C. Hsu, G.~Berie~Tarekegn, M.~Endeshaw~Getnet, K.-P. Yu, and L.-C. Tai, ``Uav-based real-time air quality monitoring and prediction using machine learning approach,'' \emph{IEEE Internet of Things Journal}, vol.~12, no.~14, pp. 27\,915--27\,928, 2025.

\bibitem{ref23}
G.~Sun, W.~Xie, D.~Niyato, H.~Du, J.~Kang, J.~Wu, S.~Sun, and P.~Zhang, ``Generative ai for advanced uav networking,'' \emph{IEEE Network}, vol.~39, no.~4, pp. 244--253, 2025.

\bibitem{ref24}
N.~Cheng, S.~Wu, X.~Wang, Z.~Yin, C.~Li, W.~Chen, and F.~Chen, ``Ai for uav-assisted iot applications: A comprehensive review,'' \emph{IEEE Internet of Things Journal}, vol.~10, no.~16, pp. 14\,438--14\,461, 2023.

\bibitem{ref25}
A.-R. Shaout, M.~Zohdy, and A.~Shaout, ``A novel approach to a generalized classification of autonomous uavs,'' in \emph{2023 24th International Arab Conference on Information Technology (ACIT)}, 2023, pp. 1--6.

\bibitem{ref26}
Y.~Gao and Q.~Jia, ``Uav swarm network service classification based on multi-target threat assessment,'' in \emph{2024 4th International Conference on Artificial Intelligence, Robotics, and Communication (ICAIRC)}, 2024, pp. 776--780.

\bibitem{ref27}
Y.~Li, Y.~Zhu, T.~Zhang, and D.~Fan, ``Joint doppler shift and channel estimation for uav mmwave system with massive ula,'' \emph{China Communications}, vol.~19, no.~4, pp. 67--82, 2022.

\bibitem{ref28}
D.~Liu, B.~Fei, W.~Bao, X.~Zhu, and X.~Li, ``Dawn: Dynamic task planning of multi-uav with two-layer optimization mechanism in uncertain environments,'' \emph{IEEE Internet of Things Journal}, vol.~11, no.~23, pp. 37\,813--37\,830, 2024.

\bibitem{ref118}
H.~Qiu, K.~Zhu, N.~C. Luong, C.~Yi, D.~Niyato, and D.~I. Kim, ``Applications of auction and mechanism design in edge computing: A survey,'' \emph{IEEE Transactions on Cognitive Communications and Networking}, vol.~8, no.~2, pp. 1034--1058, 2022.

\bibitem{ref119}
A.~u.~R. Khan, M.~Othman, S.~A. Madani, and S.~U. Khan, ``A survey of mobile cloud computing application models,'' \emph{IEEE Communications Surveys \& Tutorials}, vol.~16, no.~1, pp. 393--413, 2014.

\bibitem{ref120}
S.~Douch, M.~R. Abid, K.~Zine-Dine, D.~Bouzidi, and D.~Benhaddou, ``Edge computing technology enablers: A systematic lecture study,'' \emph{IEEE Access}, vol.~10, pp. 69\,264--69\,302, 2022.

\bibitem{ref121}
I.~Sadooghi, J.~H. Martin, T.~Li, K.~Brandstatter, K.~Maheshwari, T.~P.~P. de~Lacerda~Ruivo, G.~Garzoglio, S.~Timm, Y.~Zhao, and I.~Raicu, ``Understanding the performance and potential of cloud computing for scientific applications,'' \emph{IEEE Transactions on Cloud Computing}, vol.~5, no.~2, pp. 358--371, 2017.

\bibitem{ref122}
S.~S. Hajam and S.~A. Sofi, ``Iot-fog architectures in smart city applications: A survey,'' \emph{China Communications}, vol.~18, no.~11, pp. 117--140, 2021.

\bibitem{ref123}
J.~Pan and J.~McElhannon, ``Future edge cloud and edge computing for internet of things applications,'' \emph{IEEE Internet of Things Journal}, vol.~5, no.~1, pp. 439--449, 2018.

\bibitem{ref3}
Y.~Tian, Z.~Zhang, Y.~Yang, Z.~Chen, Z.~Yang, R.~Jin, T.~Q.~S. Quek, and K.-K. Wong, ``An edge-cloud collaboration framework for generative ai service provision with synergetic big cloud model and small edge models,'' \emph{IEEE Network}, vol.~38, no.~5, pp. 37--46, 2024.

\bibitem{ref115}
S.~Kianpisheh and T.~Taleb, ``A survey on in-network computing: Programmable data plane and technology specific applications,'' \emph{IEEE Communications Surveys \& Tutorials}, vol.~25, no.~1, pp. 701--761, 2023.

\bibitem{ref116}
N.~Derić, A.~Varasteh, A.~Blenk, and W.~Kellerer, ``Naga: A deterministic programmable network with update timing guarantees,'' \emph{IEEE Transactions on Network and Service Management}, vol.~22, no.~2, pp. 1874--1888, 2025.

\bibitem{ref68}
J.~Suo, X.~Zhang, W.~Shi, and W.~Zhou, ``E3-uav: An edge-based energy-efficient object detection system for unmanned aerial vehicles,'' \emph{IEEE Internet of Things Journal}, vol.~11, no.~3, pp. 4398--4413, 2024.

\bibitem{ref69}
C.~Hu, Y.~Deng, W.~Luo, Q.~Wei, and G.~Min, ``Towards a heterogeneous and elastic cloud service system with a correlation-based universal resource matching strategy,'' \emph{IEEE Transactions on Services Computing}, vol.~17, no.~5, pp. 2931--2944, 2024.

\bibitem{ref70}
D.~Agnew, A.~D. Aguila, and J.~Mcnair, ``Enhanced network metric prediction for machine learning-based cyber security of a software-defined uav relay network,'' \emph{IEEE Access}, vol.~12, pp. 54\,202--54\,219, 2024.

\bibitem{ref50}
K.~Shi, J.~Liu, X.~Wang, and L.~Xie, ``Joint optimization of multi-uav-assisted data collection and energy replenishment via transfer learning aided deep reinforcement learning,'' in \emph{2023 IEEE 23rd International Conference on Communication Technology (ICCT)}, 2023, pp. 967--972.

\bibitem{ref71}
C.~Li, Y.~Gan, Y.~Zhang, and Y.~Luo, ``A cooperative computation offloading strategy with on-demand deployment of multi-uavs in uav-aided mobile edge computing,'' \emph{IEEE Transactions on Network and Service Management}, vol.~21, no.~2, pp. 2095--2110, 2024.

\bibitem{ref67}
D.~Xu, X.~Tian, K.~Pham, E.~Blasch, and G.~Chen, ``Virtual network function placement for mapping sfc requests of uav-sourced video streaming in cloud networks,'' in \emph{2024 IEEE International Conference on Communications Workshops (ICC Workshops)}, 2024, pp. 523--528.

\bibitem{ref72}
Z.~Liu, Y.~Cao, P.~Gao, X.~Hua, D.~Zhang, and T.~Jiang, ``Multi-uav network assisted intelligent edge computing: Challenges and opportunities,'' \emph{China Communications}, vol.~19, no.~3, pp. 258--278, 2022.

\bibitem{ref73}
M.~T. Dabiri and M.~Hasna, ``A novel mrr-uav-based relay with optical network coding: A comparative study with optical irs and conventional uav relaying,'' \emph{IEEE Journal on Selected Areas in Communications}, vol.~43, no.~5, pp. 1607--1620, 2025.

\bibitem{ref74}
T.~S. Ahmed, A.~N. Sayed, A.~Youssef, G.~Shaker, and M.~Elbahnasawy, ``Assessing noise effects on uav classification accuracy with deep learning and fpga real-time processing: A study utilizing radar digital twins,'' \emph{IEEE Sensors Journal}, vol.~25, no.~12, pp. 22\,850--22\,862, 2025.

\bibitem{ref75}
Z.~Yang, S.~Bi, and Y.-J.~A. Zhang, ``Dynamic offloading and trajectory control for uav-enabled mobile edge computing system with energy harvesting devices,'' \emph{IEEE Transactions on Wireless Communications}, vol.~21, no.~12, pp. 10\,515--10\,528, 2022.

\bibitem{ref45}
Y.~Ding, Y.~Feng, W.~Lu, S.~Zheng, N.~Zhao, L.~Meng, A.~Nallanathan, and X.~Yang, ``Online edge learning offloading and resource management for uav-assisted mec secure communications,'' \emph{IEEE Journal of Selected Topics in Signal Processing}, vol.~17, no.~1, pp. 54--65, 2023.

\bibitem{ref44}
Y.~Zhou, X.~Liu, and Y.~Liu, ``Trade-off between radar sensing and energy consumption in integrated sensing, computing, and communication uav network,'' in \emph{2024 IEEE 24th International Conference on Communication Technology (ICCT)}, 2024, pp. 1378--1382.

\bibitem{ref47}
W.~Rafique, L.~Qi, I.~Yaqoob, M.~Imran, R.~U. Rasool, and W.~Dou, ``Complementing iot services through software defined networking and edge computing: A comprehensive survey,'' \emph{IEEE Communications Surveys \& Tutorials}, vol.~22, no.~3, pp. 1761--1804, 2020.

\bibitem{ref46}
P.~McEnroe, S.~Wang, and M.~Liyanage, ``A survey on the convergence of edge computing and ai for uavs: Opportunities and challenges,'' \emph{IEEE Internet of Things Journal}, vol.~9, no.~17, pp. 15\,435--15\,459, 2022.

\bibitem{ref64}
C.~Hu, Y.~Deng, W.~Luo, Q.~Wei, and G.~Min, ``Towards a heterogeneous and elastic cloud service system with a correlation-based universal resource matching strategy,'' \emph{IEEE Transactions on Services Computing}, vol.~17, no.~5, pp. 2931--2944, 2024.

\bibitem{ref76}
Y.~Zhang, Y.~Liu, G.~Sun, J.~Li, and A.~Wang, ``Multi-objective optimization for joint uav-agv collaborative beamforming,'' in \emph{2022 IEEE International Conference on Systems, Man, and Cybernetics (SMC)}, 2022, pp. 150--157.

\bibitem{ref65}
X.~Jiang, N.~Li, Y.~Guo, W.~Xie, and J.~Sheng, ``Multi-emitter localization via concurrent variational bayesian inference in uav-based wsn,'' \emph{IEEE Communications Letters}, vol.~25, no.~7, pp. 2255--2259, 2021.

\bibitem{ref77}
J.~Zhang, H.~Luo, X.~Chen, H.~Shen, and L.~Guo, ``Minimizing response delay in uav-assisted mobile edge computing by joint uav deployment and computation offloading,'' \emph{IEEE Transactions on Cloud Computing}, vol.~12, no.~4, pp. 1372--1386, 2024.

\bibitem{ref66}
J.~Zhou, J.~Wu, J.~Ni, Y.~Wang, Y.~Pan, and Z.~Su, ``Protecting your attention during distributed graph learning: Efficient privacy-preserving federated graph attention network,'' \emph{IEEE Transactions on Information Forensics and Security}, vol.~20, pp. 1949--1964, 2025.

\bibitem{ref78}
A.~Yao, S.~Pal, C.~Dong, X.~Li, and X.~Liu, ``A framework for user biometric privacy protection in uav delivery systems with edge computing,'' in \emph{2024 IEEE International Conference on Pervasive Computing and Communications Workshops and other Affiliated Events (PerCom Workshops)}, 2024, pp. 631--636.

\bibitem{ref79}
C.~Zhan, H.~Hu, X.~Sui, Z.~Liu, and D.~Niyato, ``Completion time and energy optimization in the uav-enabled mobile-edge computing system,'' \emph{IEEE Internet of Things Journal}, vol.~7, no.~8, pp. 7808--7822, 2020.

\bibitem{ref82}
X.~Tang, Q.~Chen, W.~Weng, B.~Liao, J.~Wang, X.~Cao, and X.~Li, ``Dnn task assignment in uav networks: A generative ai enhanced multiagent reinforcement learning approach,'' \emph{IEEE Internet of Things Journal}, vol.~12, no.~10, pp. 13\,340--13\,352, 2025.

\bibitem{ref83}
Q.~Zhang, A.~Ferdowsi, W.~Saad, and M.~Bennis, ``Distributed conditional generative adversarial networks (gans) for data-driven millimeter wave communications in uav networks,'' \emph{IEEE Transactions on Wireless Communications}, vol.~21, no.~3, pp. 1438--1452, 2022.

\bibitem{ref84}
S.~Javaid, H.~Fahim, B.~He, and N.~Saeed, ``Large language models for uavs: Current state and pathways to the future,'' \emph{IEEE Open Journal of Vehicular Technology}, vol.~5, pp. 1166--1192, 2024.

\bibitem{ref80}
J.~You, Z.~Jia, C.~Dong, Q.~Wu, and Z.~Han, ``Generative ai-enhanced cooperative mec of uavs and ground stations for unmanned surface vehicles,'' in \emph{2025 59th Annual Conference on Information Sciences and Systems (CISS)}, 2025, pp. 1--6.

\bibitem{ref85}
A.~A. Al-Bakhrani, M.~Li, M.~S. Obaidat, and G.~A. Amran, ``Moalf-uav-mec: Adaptive multiobjective optimization for uav-assisted mobile edge computing in dynamic iot environments,'' \emph{IEEE Internet of Things Journal}, vol.~12, no.~12, pp. 20\,736--20\,756, 2025.

\bibitem{ref81}
L.~Zhou, W.~Feng, Z.~Chen, T.~Ruan, S.~Leng, H.~H. Yang, Y.~Fu, and T.~Q.~S. Quek, ``Cooperative generative ai for uav-based scenarios: An intelligent cooperative framework,'' \emph{IEEE Vehicular Technology Magazine}, vol.~20, no.~2, pp. 44--52, 2025.

\bibitem{ref86}
P.~Qin, Y.~Fu, J.~Zhang, S.~Geng, J.~Liu, and X.~Zhao, ``Drl-based resource allocation and trajectory planning for noma-enabled multi-uav collaborative caching 6g network,'' \emph{IEEE Transactions on Vehicular Technology}, vol.~73, no.~6, pp. 8750--8764, 2024.

\bibitem{ref87}
Z.~Bai, J.~Shi, Z.~Li, M.~Li, and K.-C. Chen, ``Rule-guided drl for uav-assisted wireless sensor networks with no-fly zones safety,'' \emph{IEEE Transactions on Cognitive Communications and Networking}, vol.~11, no.~2, pp. 1268--1280, 2025.

\bibitem{ref88}
Z.~Dai, C.~H. Liu, R.~Han, G.~Wang, K.~K. Leung, and J.~Tang, ``Delay-sensitive energy-efficient uav crowdsensing by deep reinforcement learning,'' \emph{IEEE Transactions on Mobile Computing}, vol.~22, no.~4, pp. 2038--2052, 2023.

\bibitem{ref89}
C.~Zhang, W.~Yao, Y.~Zuo, J.~Gui, and C.~Zhang, ``Multi-objective optimization of dynamic communication network for multi-uavs system,'' \emph{IEEE Transactions on Vehicular Technology}, vol.~73, no.~3, pp. 4081--4094, 2024.

\bibitem{ref90}
A.~S. Seisa, S.~G. Satpute, B.~Lindqvist, and G.~Nikolakopoulos, ``An edge architecture oriented model predictive control scheme for an autonomous uav mission,'' in \emph{2022 IEEE 31st International Symposium on Industrial Electronics (ISIE)}, 2022, pp. 1195--1201.

\bibitem{ref91}
L.~Bertizzolo, S.~D’Oro, L.~Ferranti, L.~Bonati, E.~Demirors, Z.~Guan, T.~Melodia, and S.~Pudlewski, ``Swarmcontrol: An automated distributed control framework for self-optimizing drone networks,'' in \emph{IEEE INFOCOM 2020 - IEEE Conference on Computer Communications}, 2020, pp. 1768--1777.

\bibitem{ref43}
M.~Yin, M.~Huang, R.~Hu, and H.~Xu, ``Deep learning-assisted isac for uav detection in multipath environments with micro-doppler,'' \emph{IEEE Wireless Communications Letters}, pp. 1--1, 2025.

\bibitem{ref48}
\BIBentryALTinterwordspacing
W.~Jiang, W.~Yu, W.~Wang, and T.~Huang, ``Multi-agent reinforcement learning for joint cooperative spectrum sensing and channel access in cognitive uav networks,'' 2022. [Online]. Available: \url{https://arxiv.org/abs/2103.08181}
\BIBentrySTDinterwordspacing

\bibitem{ref49}
S.~R. Chintareddy, K.~Roach, K.~Cheung, and M.~Hashemi, ``Federated learning-based collaborative wideband spectrum sensing and scheduling for uavs in utm systems,'' \emph{IEEE Transactions on Machine Learning in Communications and Networking}, vol.~3, pp. 296--314, 2025.

\bibitem{ref92}
A.~Celik and A.~M. Eltawil, ``At the dawn of generative ai era: A tutorial-cum-survey on new frontiers in 6g wireless intelligence,'' \emph{IEEE Open Journal of the Communications Society}, vol.~5, pp. 2433--2489, 2024.

\bibitem{ref93}
\BIBentryALTinterwordspacing
X.~Wang, J.~Wang, L.~Feng, D.~Niyato, R.~Zhang, J.~Kang, Z.~Xiong, H.~Du, and S.~Mao, ``Wireless hallucination in generative ai-enabled communications: Concepts, issues, and solutions,'' 2025. [Online]. Available: \url{https://arxiv.org/abs/2503.06149}
\BIBentrySTDinterwordspacing

\bibitem{ref94}
\BIBentryALTinterwordspacing
Z.~Lyu, Y.~Gao, J.~Chen, H.~Du, J.~Xu, K.~Huang, and D.~I. Kim, ``Empowering intelligent low-altitude economy with large ai model deployment,'' 2025. [Online]. Available: \url{https://arxiv.org/abs/2505.22343}
\BIBentrySTDinterwordspacing

\bibitem{ref95}
M.~Leranso~Betalo, I.~Ullah, F.~Berhanu~Tesema, Z.~Wu, J.~Li, and X.~Bai, ``Generative ai-driven multiagent drl for task allocation in uav-assisted empd within 6g-enabled sagin networks,'' \emph{IEEE Internet of Things Journal}, vol.~12, no.~17, pp. 35\,890--35\,907, 2025.

\bibitem{ref96}
\BIBentryALTinterwordspacing
A.~Ray, ``Edgeagentx-dt: Integrating digital twins and generative ai for resilient edge intelligence in tactical networks,'' 2025. [Online]. Available: \url{https://arxiv.org/abs/2507.21196}
\BIBentrySTDinterwordspacing

\bibitem{ref97}
J.~Xia, P.~Wang, B.~Li, and Z.~Fei, ``Intelligent task offloading and collaborative computation in multi-uav-enabled mobile edge computing,'' \emph{China Communications}, vol.~19, no.~4, pp. 244--256, 2022.

\bibitem{ref98}
A.~M. Benaya, M.~S. Hassan, M.~H. Ismail, and T.~Landolsi, ``Aerial isac: A haps-assisted integrated sensing, communications and computing framework for enhanced coverage and security,'' \emph{IEEE Transactions on Green Communications and Networking}, pp. 1--1, 2025.

\bibitem{ref99}
S.~Li, Y.~Jia, F.~Yang, Q.~Qin, H.~Gao, and Y.~Zhou, ``Collaborative decision-making method for multi-uav based on multiagent reinforcement learning,'' \emph{IEEE Access}, vol.~10, pp. 91\,385--91\,396, 2022.

\bibitem{ref52}
Z.~Yang, M.~Chen, G.~Li, Y.~Yang, and Z.~Zhang, ``Secure semantic communications: Fundamentals and challenges,'' \emph{IEEE Network}, vol.~38, no.~6, pp. 513--520, 2024.

\bibitem{ref53}
X.~Yao, J.~Zheng, X.~Zheng, H.~Dai, and X.~Yang, ``Optimization of image transmission in uav-enabled semantic communication networks,'' in \emph{2023 9th International Conference on Computer and Communications (ICCC)}, 2023, pp. 647--652.

\bibitem{ref54}
Y.~Long, S.~Gong, S.~Sun, G.~C. Lee, L.~Li, and D.~Niyato, ``Lyapunov-guided deep reinforcement learning for semantic-aware aoi minimization in uav-assisted wireless networks,'' \emph{IEEE Transactions on Wireless Communications}, pp. 1--1, 2025.

\bibitem{ref56}
Z.~Chen, S.~Zhang, Y.~Tang, M.~Xu, X.~Wu, M.~Ye, Z.~Zhang, Y.~Qi, H.~Lu, and W.~Huo, ``Optimization of computing power network routing strategies based on multi-agent soft actor-critic,'' in \emph{2024 9th International Conference on Intelligent Computing and Signal Processing (ICSP)}, 2024, pp. 426--430.

\bibitem{ref57}
Y.~Qiu, F.~Chu, Y.~Lu, and L.~Gan, ``A programmable computing power network forwarding and scheduling method based on srv6,'' in \emph{2024 2nd International Conference On Mobile Internet, Cloud Computing and Information Security (MICCIS)}, 2024, pp. 204--214.

\bibitem{ref101}
A.~Varghese, A.~Heikkinen, P.~Mähönen, T.~Ojanperä, and I.~Ahmad, ``Predictive qos for cellular-connected uav communications,'' in \emph{ICC 2024 - IEEE International Conference on Communications}, 2024, pp. 3901--3906.

\bibitem{ref100}
\BIBentryALTinterwordspacing
M.~M. Rahman, ``Blockchain-assisted qos-aware routing for software-defined wide area network,'' \emph{Electronics}, vol.~14, no.~10, 2025. [Online]. Available: \url{https://www.mdpi.com/2079-9292/14/10/1949}
\BIBentrySTDinterwordspacing

\bibitem{ref58}
C.~Chen, Z.~Liu, Y.~Yu, F.~Jin, W.~Han, S.~Berretti, L.~Liu, and Q.~Pei, ``A deep-learning-based traffic classification method for 5g aerial computing networks,'' \emph{IEEE Internet of Things Journal}, vol.~12, no.~9, pp. 11\,244--11\,257, 2025.

\bibitem{ref59}
I.~Valieva and I.~Voitenko, ``Shallow neural networks for unmanned aerial vehicles data traffic classification,'' in \emph{2023 IEEE Future Networks World Forum (FNWF)}, 2023, pp. 1--6.

\bibitem{ref63}
A.~Anand, R.~S.~S. Kumar, F.~Malandra, Z.~Sun, and Z.~Guan, ``Ubspot: A universal broadband flying hotspot experimental testbed toward programmable aerial-ground wireless networks,'' in \emph{2020 IEEE Radio and Wireless Symposium (RWS)}, 2020, pp. 291--294.

\bibitem{ref102}
\BIBentryALTinterwordspacing
I.~Song, P.~Tam, S.~Kang, S.~Ros, and S.~Kim, ``Drl-based backbone sdn control methods in uav-assisted networks for computational resource efficiency,'' \emph{Electronics}, vol.~12, no.~13, 2023. [Online]. Available: \url{https://www.mdpi.com/2079-9292/12/13/2984}
\BIBentrySTDinterwordspacing

\bibitem{ref103}
Y.~Luo, W.~Ding, and B.~Zhang, ``Optimization of task scheduling and dynamic service strategy for multi-uav-enabled mobile-edge computing system,'' \emph{IEEE Transactions on Cognitive Communications and Networking}, vol.~7, no.~3, pp. 970--984, 2021.

\bibitem{ref104}
\BIBentryALTinterwordspacing
O.~Bekkouche, M.~Bagaa, and T.~Taleb, ``Toward a utm-based service orchestration for uavs in mec-nfv environment,'' 2022. [Online]. Available: \url{https://arxiv.org/abs/2201.00994}
\BIBentrySTDinterwordspacing

\bibitem{ref105}
\BIBentryALTinterwordspacing
T.~Azfar, K.~Huang, and R.~Ke, ``Enhancing disaster resilience with uav-assisted edge computing: A reinforcement learning approach to managing heterogeneous edge devices,'' 2025. [Online]. Available: \url{https://arxiv.org/abs/2501.15305}
\BIBentrySTDinterwordspacing

\bibitem{ref106}
\BIBentryALTinterwordspacing
I.~D.~I. Saeedi and A.~K.~M. Al-Qurabat, ``A comprehensive review of computation offloading in uav-assisted mobile edge computing for iot applications,'' \emph{Physical Communication}, vol.~72, p. 102810, 2025. [Online]. Available: \url{https://www.sciencedirect.com/science/article/pii/S1874490725002137}
\BIBentrySTDinterwordspacing

\bibitem{ref61}
\BIBentryALTinterwordspacing
D.~Candal‐Ventureira, F.~J. González‐Castaño, F.~Gil‐Castiñeira, and P.~Fondo‐Ferreiro, ``Is the edge really necessary for drone computing offloading? an experimental assessment in carrier‐grade 5g operator networks,'' \emph{Software: Practice and Experience}, vol.~53, no.~3, p. 579–599, Oct. 2022. [Online]. Available: \url{http://dx.doi.org/10.1002/spe.3161}
\BIBentrySTDinterwordspacing

\bibitem{ref62}
\BIBentryALTinterwordspacing
Z.~Fang, Z.~Liu, J.~Wang, S.~Hu, Y.~Guo, Y.~Deng, and Y.~Fang, ``Task-oriented communications for visual navigation with edge-aerial collaboration in low altitude economy,'' 2025. [Online]. Available: \url{https://arxiv.org/abs/2504.18317}
\BIBentrySTDinterwordspacing

\bibitem{ref107}
\BIBentryALTinterwordspacing
Q.~Chen, H.~Zhu, L.~Yang, X.~Chen, S.~Pollin, and E.~Vinogradov, ``Edge computing assisted autonomous flight for uav: Synergies between vision and communications,'' 2020. [Online]. Available: \url{https://arxiv.org/abs/2012.05517}
\BIBentrySTDinterwordspacing

\bibitem{ref108}
G.~Chen, C.~Chang, R.~Pang, T.~He, F.~Han, and J.~Pan, ``Analysis and research on key technologies of cdn based on the integration of computing and network,'' in \emph{2024 5th International Symposium on Computer Engineering and Intelligent Communications (ISCEIC)}, 2024, pp. 115--119.

\bibitem{ref109}
\BIBentryALTinterwordspacing
K.~Xu, S.~Bin, G.~Jie, Q.~Zhijin, and F.~R. Yu, ``Task-oriented image transmission for scene classification in unmanned aerial systems,'' 2021. [Online]. Available: \url{https://arxiv.org/abs/2112.10948}
\BIBentrySTDinterwordspacing

\bibitem{ref110}
M.~M.~U. Rathore, A.~Paul, A.~Ahmad, B.-W. Chen, B.~Huang, and W.~Ji, ``Real-time big data analytical architecture for remote sensing application,'' \emph{IEEE Journal of Selected Topics in Applied Earth Observations and Remote Sensing}, vol.~8, no.~10, pp. 4610--4621, 2015.

\bibitem{ref136}
T.~Wang, Y.~Lu, J.~Wang, H.-N. Dai, X.~Zheng, and W.~Jia, ``Eihdp: Edge-intelligent hierarchical dynamic pricing based on cloud-edge-client collaboration for iot systems,'' \emph{IEEE Transactions on Computers}, vol.~70, no.~8, pp. 1285--1298, 2021.

\bibitem{ref137}
S.~Kianpisheh and T.~Taleb, ``A survey on in-network computing: Programmable data plane and technology specific applications,'' \emph{IEEE Communications Surveys \& Tutorials}, vol.~25, no.~1, pp. 701--761, 2023.

\bibitem{ref124}
Y.~Liu, S.~Xie, and Y.~Zhang, ``Cooperative offloading and resource management for uav-enabled mobile edge computing in power iot system,'' \emph{IEEE Transactions on Vehicular Technology}, vol.~69, no.~10, pp. 12\,229--12\,239, 2020.

\bibitem{ref125}
J.~Tang and Y.~Zeng, ``Uav data acquisition and processing assisted by ugv-enabled mobile edge computing,'' \emph{IEEE Transactions on Industrial Informatics}, vol.~21, no.~5, pp. 3695--3704, 2025.

\bibitem{ref138}
L.~Zuo, Y.~Li, S.~Xia, and J.~Pan, ``Blockchain-based collaborative task offloading algorithm in heterogeneous edge computing networks,'' \emph{IEEE Transactions on Cognitive Communications and Networking}, pp. 1--1, 2025.

\bibitem{ref139}
Y.~Yang, Z.~Ke, X.~Jia, and L.~Yan, ``Intelligent optimization strategies for hierarchical cloud-edge computing in heterogeneous hardware ecosystems,'' in \emph{2025 4th International Conference on Electronics, Integrated Circuits and Communication Technology (EICCT)}, 2025, pp. 578--581.

\bibitem{ref126}
L.~Tan, S.~Guo, P.~Zhou, Z.~Kuang, S.~Long, and Z.~Li, ``Multi-uav-enabled collaborative edge computing: Deployment, offloading and resource optimization,'' \emph{IEEE Transactions on Intelligent Transportation Systems}, vol.~25, no.~11, pp. 18\,305--18\,320, 2024.

\bibitem{ref127}
S.~Han, X.~Liu, M.~Zhou, K.~Zhu, L.~Zhao, A.~Albeshri, and A.~Abusorrah, ``Joint association, deployment and flight trajectory optimization for multi-uav-enabled large-scale mobile edge computing,'' \emph{IEEE Transactions on Mobile Computing}, vol.~23, no.~12, pp. 13\,207--13\,221, 2024.

\bibitem{ref140}
Z.~Guo, J.~Wang, S.~Liu, J.~Ren, Y.~Xu, and C.~Yao, ``Congestion-aware critical gradient scheduling for distributed machine learning in data center networks,'' \emph{IEEE Transactions on Cloud Computing}, vol.~11, no.~3, pp. 2296--2311, 2023.

\bibitem{ref141}
K.~Liu, J.~Peng, J.~Wang, B.~Yu, Z.~Liao, Z.~Huang, and J.~Pan, ``A learning-based data placement framework for low latency in data center networks,'' \emph{IEEE Transactions on Cloud Computing}, vol.~10, no.~1, pp. 146--157, 2022.

\bibitem{ref128}
\BIBentryALTinterwordspacing
R.~Arranz, D.~Carramiñana, G.~d. Miguel, J.~A. Besada, and A.~M. Bernardos, ``Application of deep reinforcement learning to uav swarming for ground surveillance,'' \emph{Sensors}, vol.~23, no.~21, p. 8766, Oct. 2023. [Online]. Available: \url{http://dx.doi.org/10.3390/s23218766}
\BIBentrySTDinterwordspacing

\bibitem{ref129}
\BIBentryALTinterwordspacing
J.-M. Fortin, O.~Gamache, W.~Fecteau, E.~Daum, W.~Larrivée-Hardy, F.~Pomerleau, and P.~Giguère, ``Uav-assisted self-supervised terrain awareness for off-road navigation,'' 2025. [Online]. Available: \url{https://arxiv.org/abs/2409.18253}
\BIBentrySTDinterwordspacing

\bibitem{ref142}
J.~Liu, P.~Xun, and B.~Wang, ``Relevant backtracking: An efficient telemetry data collection method for data center networks,'' in \emph{2023 15th International Conference on Communication Software and Networks (ICCSN)}, 2023, pp. 100--107.

\bibitem{ref130}
T.~Sun, X.~Wang, X.~Ye, and B.~Han, ``Stitch-able split learning assisted multi-uav systems,'' \emph{IEEE Open Journal of the Computer Society}, vol.~5, pp. 418--429, 2024.

\bibitem{ref131}
\BIBentryALTinterwordspacing
H.~Li, P.~Gong, S.~Li, W.~Wang, Y.~Liu, X.~Gao, D.~O. Wu, D.~K. Kim, G.~Zhang, and J.~Zhang, ``Collaborative federated learning of unmanned aerial vehicles in space–air–ground integrated network,'' \emph{Space: Science \&amp; Technology}, vol.~5, p. 0264, 2025. [Online]. Available: \url{https://spj.science.org/doi/abs/10.34133/space.0264}
\BIBentrySTDinterwordspacing

\bibitem{ref143}
A.~Ghorab, M.~Abuibaid, and M.~St-Hilaire, ``Sdn-based service discovery and assignment framework to preserve service availability in telco-based multi-access edge computing,'' in \emph{2022 IEEE 6th International Conference on Fog and Edge Computing (ICFEC)}, 2022, pp. 100--104.

\bibitem{ref132}
I.~Cardei, M.~Cardei, and R.~Papa, ``Uav-enabled data gathering in wireless sensor networks,'' in \emph{2018 IEEE 37th International Performance Computing and Communications Conference (IPCCC)}, 2018, pp. 1--8.

\bibitem{ref133}
M.~R. Ramzan, M.~Naeem, M.~Altaf, and W.~Ejaz, ``Multicriterion resource management in energy-harvested cooperative uav-enabled iot networks,'' \emph{IEEE Internet of Things Journal}, vol.~9, no.~4, pp. 2944--2959, 2022.

\bibitem{ref144}
D.~Shen, X.~Fang, W.~Feng, Y.~Chen, and Z.~Pang, ``Joint communication bandwidth and computing frequency allocation for control-oriented uav-robot rescue systems,'' in \emph{2024 IEEE 24th International Conference on Communication Technology (ICCT)}, 2024, pp. 235--239.

\bibitem{ref134}
H.~Qi, Y.~Li, Y.~Xu, and T.~Q.~S. Quek, ``Uav-assisted mobile crowdsensing systems empowered by wireless power transfer and adaptive compression techniques,'' \emph{IEEE Wireless Communications Letters}, vol.~13, no.~9, pp. 2487--2491, 2024.

\bibitem{ref135}
\BIBentryALTinterwordspacing
X.~Hu, H.~Zhao, D.~He, and W.~Zhang, ``Secure communication and resource allocation in double-ris cooperative-aided uav-mec networks,'' \emph{Drones}, vol.~9, no.~8, 2025. [Online]. Available: \url{https://www.mdpi.com/2504-446X/9/8/587}
\BIBentrySTDinterwordspacing

\bibitem{ref149}
Z.~Chen, S.~Zhang, Y.~Tang, M.~Xu, X.~Wu, M.~Ye, Z.~Zhang, Y.~Qi, H.~Lu, and W.~Huo, ``Optimization of computing power network routing strategies based on multi-agent soft actor-critic,'' in \emph{2024 9th International Conference on Intelligent Computing and Signal Processing (ICSP)}, 2024, pp. 426--430.

\bibitem{ref145}
X.~Zhao, H.~Yang, and M.~Li, ``Graph reinforcement learning based multi-hotspot region uav dynamic scheduling in mobile edge computing,'' in \emph{2024 IEEE Wireless Communications and Networking Conference (WCNC)}, 2024, pp. 1--6.

\bibitem{ref146}
Z.~Zhao, L.~Pacheco, H.~Santos, M.~Liu, A.~D. Maio, D.~Rosário, E.~Cerqueira, T.~Braun, and X.~Cao, ``Predictive uav base station deployment and service offloading with distributed edge learning,'' \emph{IEEE Transactions on Network and Service Management}, vol.~18, no.~4, pp. 3955--3972, 2021.

\bibitem{ref150}
J.~Yang, Q.~Wu, Z.~Feng, Z.~Zhou, D.~Guo, and X.~Chen, ``Quality-of-service aware llm routing for edge computing with multiple experts,'' \emph{IEEE Transactions on Mobile Computing}, pp. 1--15, 2025.

\bibitem{ref147}
\BIBentryALTinterwordspacing
O.~S. Oubbati, A.~Lakas, F.~Zhou, M.~Güneş, N.~Lagraa, and M.~B. Yagoubi, ``Intelligent uav-assisted routing protocol for urban vanets,'' \emph{Computer Communications}, vol. 107, pp. 93--111, 2017. [Online]. Available: \url{https://www.sciencedirect.com/science/article/pii/S0140366417303948}
\BIBentrySTDinterwordspacing

\bibitem{ref148}
B.~N. Getu and A.~Alnoman, ``Wireless routeing scheme for uav-assisted access networks based on the link path loss,'' \emph{Journal of Information and Telecommunication}, vol.~9, no.~2, pp. 279--293, 2025.

\bibitem{ref155}
N.~T. Hoa, C.~T.~T. Hai, H.~L. Hung, N.~Cong~Luong, and D.~Niyato, ``Joint edge computing and semantic communication in uav-enabled networks,'' \emph{IEEE Communications Letters}, vol.~29, no.~1, pp. 80--84, 2025.

\bibitem{ref156}
J.~Wang, Y.~Liu, S.~Niu, and H.~Song, ``Integration of software defined radios and software defined networking towards reinforcement learning enabled unmanned aerial vehicle networks,'' in \emph{2019 IEEE International Conference on Industrial Internet (ICII)}, 2019, pp. 44--49.

\bibitem{ref151}
Z.~Xu, I.~Petrunin, and A.~Tsourdos, ``Dynamic spectrum management with network function virtualization for uav communication,'' \emph{Journal of Intelligent \& Robotic Systems}, vol. 101, no.~40, 2021.

\bibitem{ref152}
T.~Zhou, C.~Liang, and Q.~Chen, ``Joint caching and computing of software-defined space-air-ground integrated networks for video streaming service improvement,'' in \emph{2022 IEEE 96th Vehicular Technology Conference (VTC2022-Fall)}, 2022, pp. 1--5.

\bibitem{ref153}
\BIBentryALTinterwordspacing
N.~Zhang, S.~Zhang, P.~Yang, O.~Alhussein, W.~Zhuang, and X.~Shen, ``Software defined space-air-ground integrated vehicular networks: Challenges and solutions,'' 2017. [Online]. Available: \url{https://arxiv.org/abs/1703.02664}
\BIBentrySTDinterwordspacing

\bibitem{ref154}
C.~Li, Y.~Gan, Y.~Zhang, and Y.~Luo, ``A cooperative computation offloading strategy with on-demand deployment of multi-uavs in uav-aided mobile edge computing,'' \emph{IEEE Transactions on Network and Service Management}, vol.~21, no.~2, pp. 2095--2110, 2024.

\bibitem{ref157}
L.~Lei, G.~Shen, L.~Zhang, and Z.~Li, ``Toward intelligent cooperation of uav swarms: When machine learning meets digital twin,'' \emph{IEEE Network}, vol.~35, no.~1, pp. 386--392, 2021.

\bibitem{ref158}
X.~Huang, Y.~Zhang, Y.~Qi, C.~Huang, and M.~S. Hossain, ``Energy-efficient uav scheduling and probabilistic task offloading for digital twin-empowered consumer electronics industry,'' \emph{IEEE Transactions on Consumer Electronics}, vol.~70, no.~1, pp. 2145--2154, 2024.

\bibitem{ref160}
N.~K. Bairwa, G.~Jain, S.~Umrao, and P.~Ponde, ``Impact of ai driven predictive privacy mechanisms in uav integrated v2x 6g networks,'' in \emph{2024 International Conference on Emerging Technologies and Innovation for Sustainability (EmergIN)}, 2024, pp. 135--139.

\bibitem{ref161}
M.~Ibrahim and M.~I.~A. Zahed, ``Uav-integrated blockchain-enabled optimal video caching and transcoding,'' in \emph{2024 27th International Conference on Computer and Information Technology (ICCIT)}, 2024, pp. 1081--1086.

\bibitem{ref159}
R.~Wang, X.~Liu, L.~Xie, Y.~Liu, Z.~Su, D.~Liu, and H.~Zhang, ``Privacy-preserving incentive scheme design for uav-enabled federated learning,'' in \emph{2024 IEEE Wireless Communications and Networking Conference}, 2024, pp. 1--6.

\bibitem{ref162}
M.~Ibrahim and M.~I.~A. Zahed, ``Uav-integrated blockchain-enabled optimal video caching and transcoding,'' in \emph{2024 27th International Conference on Computer and Information Technology (ICCIT)}, 2024, pp. 1081--1086.

\bibitem{ref163}
Y.~Luo, X.~Xing, and G.~Fan, ``A heterogeneous multi-uav task allocation method based on extended cbba,'' in \emph{2024 36th Chinese Control and Decision Conference (CCDC)}, 2024, pp. 431--436.

\bibitem{ref164}
R.~Samy, H.-C. Yang, T.~Rakia, and M.-S. Alouini, ``Space-air-ground fso networks for high-throughput satellite communications,'' \emph{IEEE Communications Magazine}, vol.~61, no.~3, pp. 82--87, 2023.

\bibitem{ref165}
G.~Zeng, Y.~Zhan, and H.~Xie, ``Channel allocation for mega leo satellite constellations in the meo–leo networked telemetry system,'' \emph{IEEE Internet of Things Journal}, vol.~10, no.~3, pp. 2545--2556, 2023.

\bibitem{ref166}
A.~A. Kılıç and F.~A. Yapar, ``Antipodal vivaldi antenna design for 6g high altitude platform system (haps),'' in \emph{2023 31st Signal Processing and Communications Applications Conference (SIU)}, 2023, pp. 1--4.

\end{thebibliography}
\bibliographystyle{IEEEtran}

\end{document}